\documentclass[journal]{IEEEtran}
\usepackage{setspace}
 \usepackage{cite}
 \usepackage{calligra}
 \usepackage{latexsym}
 \usepackage{times}
 \usepackage{graphicx}
 \usepackage{color}
\usepackage{algorithm}
\usepackage{algorithmic}
 \usepackage{amssymb,amsmath}
 \usepackage{multirow}
 \usepackage{colortbl} 
\usepackage{enumerate}
\usepackage{graphicx}
\usepackage{caption}
\usepackage{subcaption}
\usepackage{mathrsfs}
\usepackage{amsbsy}
\usepackage{xcolor, colortbl}
\usepackage{mathtools}
\usepackage[utf8]{inputenc}
\usepackage{setspace}
 \usepackage{calligra}
 \usepackage{latexsym}
 \usepackage{times}
 \usepackage{graphicx}
\usepackage{algorithm}
\usepackage{algorithmic}
 \usepackage{amssymb,amsmath}
 \usepackage{multirow}
 \usepackage{mathtools}
\usepackage{commath}
\usepackage{tikz}
\usepackage{tabularx}
\usepackage{enumitem, kantlipsum}
\usetikzlibrary{positioning,decorations.markings,calc}

\usetikzlibrary{shapes}

\usetikzlibrary{automata,positioning}
\usetikzlibrary{backgrounds,automata}
\usetikzlibrary{arrows, arrows.meta}
\usetikzlibrary{positioning,fit,calc}
\definecolor{LightCyan}{rgb}{0.9,1,1}
\definecolor{Gray}{gray}{0.85}

\newcolumntype{L}{>{\raggedright\arraybackslash}X}

\begin{document}
\author{Satyam Dwivedi, Ritesh Shreevastav, Florent Munier, Johannes Nygren, Iana Siomina, Yazid Lyazidi,  Deep Shrestha, Gustav Lindmark, Per Ernström, Erik Stare, Sara M. Razavi, Siva Muruganathan, Gino Masini, Åke Busin, Fredrik Gunnarsson }
\title{Positioning in 5G networks}

\maketitle 

\begin{abstract}

In this paper we describe the recent 3GPP Release 16 specification for positioning in 5G networks. It specifies positioning signals, measurements, procedures, and architecture to meet requirements from a plethora of regulatory, commercial and industrial use cases. 5G thereby significantly extends positioning capabilities compared to what was possible with LTE. The indicative positioning performance is evaluated in agreed representative 3GPP simulation scenarios, showing a 90 percentile accuracy of a few meters down to a few decimeters depending on scenarios 
and assumptions.
  
\end{abstract}

\section{Introduction}

5G and digitalization are often closely related, and with position information being central in digitalization, the introduction of 5G positioning is a vital step. In Release 16, LTE positioning feature is extended to accommodate enablers of 5G such as wideband signals, higher frequencies, multiple antennas, low latency and flexible architecture. This article presents methods, architecture, procedures, signals and measurements for 5G positioning recently specified in Release 16.

5G is designed to address requirements and needs of industrial verticals.  The 3rd Generation Partnership Project
(3GPP) Service and System Aspects (SA) specifications and technical reports on positioning requirements across industry verticals suggests accuracy target from tens of meters for emergency calls, to a few decimeters within indoor factory and one decimeter for vehicle-to-everything (V2X) use cases \cite{22.261, 22.804}. The required user equipment (UE) positioning accuracy for LTE was motivated by regulatory positioning requirements set by FCC \cite{8030544, 8377447}. For 5G, the requirements are motivated by commercial use cases. A few key enablers for precise positioning in 5G include mmWave frequency bands, which enables wideband signals, beamforming and precise angle estimation with multiple antennas \cite{8258595, rahman2020enabling}. However, most features are also available at low and mid frequency ranges. In this paper, we evaluate positioning performance in different frequency ranges.
The Global Navigation Satellite System (GNSS) is an example of a successful positioning technology but restricted to outdoor scenarios. The 5G will bring high accuracy positioning to indoor scenarios while also providing  better positioning accuracy outdoors than possible with LTE or GNSS alone \cite{9053157}.


It is expected that many features useful for positioning can be extracted from the specified elements than only the intended features during the standardization. Such as, fingerprinting, radio network optimization, soft information extraction et cetera \cite{6924849, 8827486}.  It is also possible to use signals and measurements defined for mobility and radio resource management for positioning. Such enablers can be suitable for positioning in the contexts of radio network management and analytics. The full positioning potential of the offerings from the standard is possible with a UE and the network cooperating with each other. Elements of specifications can still be useful when such cooperations are limited.

Release 16 specifies positioning signals and measurements for the 5G New Radio (NR). In subsequent sections we explain the new 5G positioning improvements in architecture, signals and measurements.  Moreover we also demonstrate possible positioning accuracy with these new capabilities.


\begin{figure}
    \begin{subfigure}[t]{0.4\linewidth}
\begin{tikzpicture}
  \node at (-0.15,.5) [rectangle,draw=black!50,fill=gray!20] (A){UE};

   \node (dep) at (2, 2.75) { \small{NG-RAN} };
     \draw [draw=black!50, dashed] (1.0,0) rectangle (3,3);

    \node (dep2) at (2.25, 2.2) { \small{ng-eNB} }; 
    \draw [draw=black, fill=magenta!80, opacity=0.1] (1.25, 1.75) rectangle (2.75, 2.5);
    \draw [draw=black!50, dashed] (1.3, 2.16) rectangle (1.7, 2.4);
    \draw [draw=black!50, dashed] (1.3, 1.85) rectangle (1.7, 2.09);
    \node (dep2) at (1.45, 2.27) { \tiny{TP} };
    \node (dep2) at (1.45, 1.97) { \tiny{TP} };

    \node (gNB) at (2.2, 0.63) { \small{gNB} };

  \draw [draw=black, fill=cyan!80, opacity=0.1] (1.25, 0.25) rectangle (2.75, 1);
    \draw [draw=black!50, dashed] (1.3, 0.67) rectangle (1.7, 0.9);
    \draw [draw=black!50, dashed] (1.3, 0.36) rectangle (1.7, 0.59);
    \node (dep2) at (1.49, 0.81) { \tiny{TRP} };
    \node (dep2) at (1.49, 0.46) { \tiny{TRP} };

    \draw[<-, blue, thick, dash pattern= on 3pt off 5pt] (0.25,0.5) -- (1.2,0.5);
    \draw[->, black!40!green, thick, dash pattern= on 3pt off 5pt, dash phase=4pt] (0.25,0.5) -- (1.2,0.5);

    \draw[<-, blue, thick, dash pattern= on 3pt off 5pt] (2.75,0.5) -- (3.6,0.5);
    \draw[->, black!40!red, thick, dash pattern= on 3pt off 5pt, dash phase=4pt] (2.75,0.5) -- (3.6,0.5);

    \node at (0.65, 0.65) { \tiny{NR-Uu} };
    \node at (3.15, 0.65) { \tiny{NG-C} };

    \node at (0.6, 1.5) [rotate=56]{ \tiny{LTE-Uu} };

    \draw[<->, black!40!green, thick, dash pattern= on 4pt off 5pt] (0.25,0.7) -- (1.2,2.1);

    \draw[<->,black!40!red, thick, dash pattern= on 4pt off 5pt] (2.8,2.1) -- (3.6,0.7);

    \node at (3.3, 1.5) [rotate=-56]{ \tiny{NG-C} };

    \draw[ ->, black!40!red, thick, dash pattern= on 4pt off 5pt] (4.1,1.5) -- (4.1,2.3) ;
    \draw[ <-, black!40!red, thick, dash pattern= on 4pt off 5pt] (4.1, 0.75) -- (4.1,1.3) ;
    \node at (4.1, 1.4) { \tiny{NLs} };

\node at ( 4.1,0.5) [rectangle,draw=black!50,fill=gray!20]  (B){AMF};
\node at ( 4.1,2.5) [rectangle,draw=black!50,fill=gray!20]  (B){LMF};



    \draw[ dashed, thick, black!40!red] (-0.3,2.7) -- (0.7,2.7) ;
    \draw[ dashed, thick, black!40!green] (-0.3, 2.3) -- (0.7,2.3) ;
    \draw[ dashed, thick, blue] (-0.3,1.9) -- (0.7,1.9) ;

    \node at (0.2, 2.85) { \tiny{NRPPa} };
    \node at (0.2, 2.45) { \tiny{RRC} };
    \node at (0.2, 2.05) { \tiny{LPP} };

\tikzset{twisted arrow/.style={%
/utils/exec=\tikzset{twisted arrow/.cd,#1},
decorate,decoration={markings,%
mark=at position 0 with {\coordinate (ta-base) at (0,0);},
mark=at position 0.02 with {\coordinate (ta-next) at (0,0);},
mark=at position 0.5 with {\coordinate (ta-mid) at (0,0);
\pgftransformreset
\filldraw[tacolor!80!black] let \p1=($(ta-next)-(ta-base)$),
    \n1={atan2(\y1,\x1)} in 
 ($(ta-base)+(\n1+90:\pgfkeysvalueof{/tikz/twisted arrow/width}/2)$) 
    to[bend left=\pgfkeysvalueof{/tikz/twisted arrow/bend angle}/2] (ta-mid) 
    to[bend right=\pgfkeysvalueof{/tikz/twisted arrow/bend angle}/2] 
 ($(ta-base)+(\n1-90:\pgfkeysvalueof{/tikz/twisted arrow/width}/2)$) 
-- cycle;},
mark=at position 0.97 with {\coordinate (ta-near-end) at (0,0);},
mark=at position 0.99 with {\coordinate (ta-end) at (0,0);
\pgftransformreset
\filldraw[tacolor] let \p1=($(ta-end)-(ta-near-end)$),
    \n1={atan2(\y1,\x1)} in 
(ta-mid) to[bend left=\pgfkeysvalueof{/tikz/twisted arrow/bend angle}/2] 
($(ta-end)+(\n1+180:\pgfkeysvalueof{/tikz/twisted arrow/head length})
+(\n1+90:\pgfkeysvalueof{/tikz/twisted arrow/width}/2)$)
-- ($(ta-end)+(\n1+180:\pgfkeysvalueof{/tikz/twisted arrow/head length})
+(\n1+90:\pgfkeysvalueof{/tikz/twisted arrow/head width}/2)$)
-- (ta-end) 
-- ($(ta-end)+(\n1+180:\pgfkeysvalueof{/tikz/twisted arrow/head length})
+(\n1-90:\pgfkeysvalueof{/tikz/twisted arrow/head width}/2)$)
--
($(ta-end)+(\n1+180:\pgfkeysvalueof{/tikz/twisted arrow/head length})
+(\n1-90:\pgfkeysvalueof{/tikz/twisted arrow/width}/2)$)
to[bend right=\pgfkeysvalueof{/tikz/twisted arrow/bend angle}/2] cycle;
}}},twisted arrow/.cd,width/.initial=6pt,
head length/.initial=6pt,
head width/.initial=10pt,
bend angle/.initial=45,
color/.code=\colorlet{tacolor}{#1},
color=gray}

\draw[twisted arrow={color=cyan!20}] (2,0.4) to[bend right=35] (4.9,0.1);
\end{tikzpicture}
    \end{subfigure}\hfill
    \begin{subfigure}[t]{0.4\linewidth}
\begin{tikzpicture}
  \node (dep) at (0.6, 2.75) { \small{gNB} };
  \draw [draw=black!50, fill=cyan!80, opacity=0.1] (0.2,0.3) rectangle (3.6,3);

 \node (dep2) at (2, 2.5) { \small{gNB-CU} };
  \draw [draw=black!50] (1,2.2) rectangle (3,2.8);

 \node (dep2) at (2, 2.5) { \small{gNB-CU} };
  \draw [draw=black!50] (0.5,0.5) rectangle (1.8,1.5);

  \draw [draw=black!50] (2.2,0.5) rectangle (3.5,1.5);

  \node (dep3) at (1.18, 1.3) { \small{gNB-DU} };
  \node (dep3) at (2.88, 1.3) { \small{gNB-DU} };
    
    \draw [draw=black!50] (0.55,0.55) rectangle (1.75,1.1);

   \draw [draw=black!50, dashed] (0.6, 0.8) rectangle (1.15, 1.05);
   \draw [draw=black!50, dashed] (1.15, 0.8) rectangle (1.7, 1.05);
  \node (dep3) at (1.15, 0.67) { \tiny{TRP} };
  \node (dep3) at (.9, 0.95) { \tiny{TP} };
  \node (dep3) at (1.45, 0.95) { \tiny{RP} };

    \draw [draw=black!50] (2.25,0.55) rectangle (3.45,1.1);

   \draw [draw=black!50, dashed] (2.3, 0.8) rectangle (2.85, 1.05);
   \draw [draw=black!50, dashed] (2.85, 0.8) rectangle (3.4, 1.05);
  \node (dep3) at (2.85, 0.67) { \tiny{TRP} };
  \node (dep3) at (2.6, 0.95) { \tiny{TP} };
  \node (dep3) at (3.15, 0.95) { \tiny{RP} };
  
    \draw[<->] (1.15,1.5) -- (1.95,2.2);
    \draw[<->] (2.85,1.5) -- (2.05,2.2);

    \node at (2.65, 1.85) [rotate=-45]{ \tiny{F1-C} };
    \node at (1.4, 1.9) [rotate=45]{ \tiny{F1-C} };

      \draw[ ->] (1.5,3.2) -- (1.5,2.8) ;
    \draw[->] (2.5, 3.2) -- (2.5,2.8) ;
    \node at (1.5, 3.3) { \tiny{NG-C} };
    \node at (2.5, 3.3) { \tiny{Xn-C} };

      \draw[ ->] (1.18,0.5) -- (1.18,0.1) ;
    \draw[->] (2.88, 0.5) -- (2.88,0.1) ;

    \node at (1.18, 0) { \tiny{NR-Uu} };
    \node at (2.88, 0) { \tiny{NR-Uu} };

\end{tikzpicture}
    \label{fig:SA}
    \end{subfigure}
    \caption{UE positioning architecture and logical protocols among different entities applicable to NG-RAN. Protocols between entities are shown using colored dashed lines.}
\label{5garch}
\end{figure}
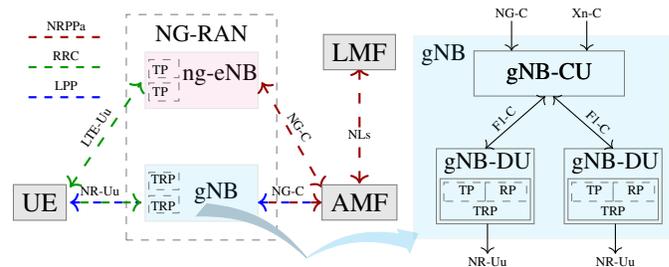

 \section{Positioning architecture and signaling protocols}
  
 The 5G positioning architecture is derived from the 4G positioning architecture, with added modifications that are consequent of the new logical nodes that are introduced to the 5G Core Network (5GC).
 Figure \ref{5garch} shows the Release 16 positioning architecture for Next Generation Radio Access Network (NG-RAN) that is applicable for positioning a UE with NR gNB Transmission Reception Points (TRPs) or LTE ng-eNB access with Transmission Points(TPs). Figure \ref{5garch} also shows the Rel-16 NG-RAN split positioning architecture. In the gNB functional split, the gNB-Central Unit (CU) and gNB Distributed Units (DU) communicate via F1 interface. As shown in the figure, the gNB-CU terminates the connection with the 5GC and can be connected to one or multiple gNB-DU, which hosts the TP/RP/TRP. Figure \ref{5garch} shows signaling protocols among different entities in the 5G positioning architecture. The gNB(gNB-CU)/ng-eNB exchanges the necessary positioning information and measurements with the Location Management Function (LMF) which is in the 5GC, via the NR Positioning Protocol Annex (NRPPa) protocol \cite{38.455}. In 4G, the positioning support between a UE and the location server is handled by the LTE Positioning Protocol (LPP). This protocol has been extended to also support 5G positioning between a UE and LMF \cite{38.305}. Whereas, the UE receives necessary radio configuration information from NG-RAN node over Radio Resource Control (RRC) via NR-Uu or LTE-Uu interface. Reusing the LPP protocol also for 5G enables extensions of both 4G and 5G in the common protocol. Both the NRPPa and the LPP protocols are transported over the control plane of the NG interface (NG-C) via Access Mobility Function (AMF)\cite{38.305}.

5G provides not only enablers for precise positioning as such, but also introduces some new positioning methods. Positioning based on multi-cell round trip time (multi-RTT) measurements, multiple antenna beam measurements to enable downlink angle of departure (DL-AoD) and uplink angle of arrival (UL-AoA) estimates has been introduced as new concepts. While the multi-RTT positioning method is robust against network time synchronization errors, angle based methods are more relevant with usage of mmWave  and multiple antennas in 5G NR. 
For multi-RTT LMF initiates the procedure whereby multiple TRPs and a UE perform the gNB Rx-Tx and UE Rx-Tx measurements respectively.
For multi-RTT, gNBs and UEs transmit  downlink positioning reference signal (DL-PRS)  and uplink sounding reference signal (UL-SRS) respectively. gNB configures UL-SRS to the UE using RRC protocol. Whereas, LMF provides the DL-PRS configuration using LPP to the UE. The UE reports the measurement results using LPP to the LMF and gNB reports the measurements using NRPPa to the LMF for UE location estimation.

For DL-AoD, UE provides the DL-PRS beam Received Signal Received Power (RSRP) measurements to LMF over LPP. The gNB provides the beam azimuth and elevation angular information to LMF over NRPPa.
In the UL AoA positioning method, the UE position is estimated based on UL SRS AoA measurements taken at different TRPs. TRPs report AoA measurements to LMF over NRPPa. Using angle informations, either AoD or AoA,  along with other configuration and deployment informations such as TRP co-ordinates and beam configuration details LMF estimates the UE location.

The Release 16 5G positioning specifications also include
NR broadcast of positioning assistance data such as for Global Navigation Satellite Systems- Real Time Kinematics
(GNSS-RTK). It enables assistance data to be either broadcasted as introduced for LTE in Relase 15 or made available on demand. With the new on-demand System Information (SI) procedure, a UE may request positioning System Information Blocks (posSIBs) by means of an on-demand SI
request (random access procedure message 1 or 3) in RRC Idle/Inactivate states and using on-demand connected mode procedure while in RRC
Connected mode.

Release 16 also enhances the scope of GNSS RTK AD with support for spatial atmospheric delay models, leveraged by the models defined for Quasi-Zenith Satellite System (QZSS). These models enables the devices to compensate for the atmospheric delays of the satellite signals.



\section{Positioning specific signals}


\begin{figure}
      \begin{subfigure}[t]{0.45\linewidth}
      \begin{tikzpicture}

    \draw [->] (0,0) -- (4.35,0);
    \draw (0,0.3) -- (3.9,0.3);
    \draw (0,0.6) -- (3.9,0.6);
    \draw (0,0.9) -- (3.9,0.9);
    \draw (0,1.2) -- (3.9,1.2);
    \draw (0,1.5) -- (3.9,1.5);
    \draw (0,1.8) -- (3.9,1.8);
    \draw (0,2.1) -- (3.9,2.1);
    \draw (0,2.4) -- (3.9,2.4);
    \draw (0,2.7) -- (3.9,2.7);
    \draw (0,3) -- (3.9,3);
    \draw (0,3.3) -- (3.9,3.3);
    \draw (0,3.6) -- (4.2,3.6);

    \draw [->] (0,0) -- (0,3.8);
    \draw (0.3,0) -- (0.3, 3.6);
    \draw (0.6, 0) -- (0.6,3.6);
    \draw (0.9,0) -- (0.9,3.6);
    \draw (1.2,0) -- (1.2,3.6);
    \draw (1.5,0) -- (1.5,3.6);
    \draw (1.8,0) -- (1.8,3.6);
    \draw (2.1,0) -- (2.1,3.6);
    \draw (2.4,0) -- (2.4,3.6);
    \draw (2.7,0) -- (2.7,3.6);
    \draw (3.0,0) -- (3.0,3.6);
    \draw (3.3,0) -- (3.3,3.6);
    \draw (3.6,0) -- (3.6,3.6);
    \draw (3.9,0) -- (3.9,3.6);
    \draw (4.2,0) -- (4.2,3.6);

 \node at (4.05, 1.8)[rotate=90] { \small{One PRB} };
    \draw [<-] (4.05,0) -- (4.05, 1);
    \draw [->] (4.05,2.6) -- (4.05,3.6);

 \node at (4.3, 0.25) {$t$ };
 \node at (0.2, 3.8){\small{$f$}};

   \draw [draw=black, fill=red] (0,0) rectangle (0.3, 0.3);
   \draw [draw=black, fill=yellow] (0,0.6) rectangle (0.3, 0.9);
   \draw [draw=black, fill=green] (0,1.2) rectangle (0.3, 1.5); 
   \draw [draw=black, fill=red] (0,1.8) rectangle (0.3, 2.1);
   \draw [draw=black, fill=yellow] (0,2.4) rectangle (0.3, 2.7);
   \draw [draw=black, fill=green] (0,3) rectangle (0.3, 3.3);

   \draw [draw=black, fill=red] (0.3,0.9) rectangle (0.6, 1.2);
   \draw [draw=black, fill=yellow] (0.3,1.5) rectangle (0.6, 1.8);
   \draw [draw=black, fill=green] (0.3,2.1) rectangle (0.6, 2.4); 
   \draw [draw=black, fill=red] (0.3,2.7) rectangle (0.6, 3);
   \draw [draw=black, fill=yellow] (0.3,3.3) rectangle (0.6, 3.6);
   \draw [draw=black, fill=green] (0.3,0.3) rectangle (0.6, 0.6);

   \draw [draw=black, fill=red] (0.6,0.3) rectangle (0.9, 0.6); 
   \draw [draw=black, fill=yellow] (0.6,0.9) rectangle (0.9, 1.2);
   \draw [draw=black, fill=green] (0.6,1.5) rectangle (0.9, 1.8); 
   \draw [draw=black, fill=red] (0.6,2.1) rectangle (0.9, 2.4);
   \draw [draw=black, fill=yellow] (0.6,2.7) rectangle (0.9, 3);
   \draw [draw=black, fill=green] (0.6,3.3) rectangle (0.9, 3.6);
    
   \draw [draw=black, fill=red] (0.9,1.2) rectangle (1.2, 1.5);
   \draw [draw=black, fill=yellow] (0.9,1.8) rectangle (1.2, 2.1);
   \draw [draw=black, fill=green] (0.9,2.4) rectangle (1.2, 2.7); 
   \draw [draw=black, fill=red] (0.9,3) rectangle (1.2, 3.3);
   \draw [draw=black, fill=yellow] (0.9,0) rectangle (1.2, 0.3);
   \draw [draw=black, fill=green] (0.9,0.6) rectangle (1.2, 0.9);

  \draw [draw=black, fill=red] (1.2,2.4) rectangle (1.5, 2.7);
   \draw [draw=black, fill=yellow] (1.2,3) rectangle (1.5, 3.3);
   \draw [draw=black, fill=green] (1.2,0) rectangle (1.5, 0.3); 
   \draw [draw=black, fill=red] (1.2,0.6) rectangle (1.5, 0.9);
   \draw [draw=black, fill=yellow] (1.2,1.2) rectangle (1.5, 1.5);
   \draw [draw=black, fill=green] (1.2,1.8) rectangle (1.5, 2.1);

   \draw [draw=black, fill=red] (1.5,1.5) rectangle (1.8, 1.8);
   \draw [draw=black, fill=yellow] (1.5,2.1) rectangle (1.8, 2.4);
   \draw [draw=black, fill=green] (1.5,2.7) rectangle (1.8, 3); 
   \draw [draw=black, fill=red] (1.5,3.3) rectangle (1.8, 3.6);
   \draw [draw=black, fill=yellow] (1.5,0.3) rectangle (1.8, 0.6);
   \draw [draw=black, fill=green] (1.5,0.9) rectangle (1.8, 1.2);

   \draw [draw=black, fill=red] (1.8,0) rectangle (2.1, 0.3);
   \draw [draw=black, fill=yellow] (1.8,0.6) rectangle (2.1, 0.9);
   \draw [draw=black, fill=green] (1.8,1.2) rectangle (2.1, 1.5); 
   \draw [draw=black, fill=red] (1.8,1.8) rectangle (2.1, 2.1);
   \draw [draw=black, fill=yellow] (1.8,2.4) rectangle (2.1, 2.7);
   \draw [draw=black, fill=green] (1.8,3) rectangle (2.1, 3.3);

   \draw [draw=black, fill=red] (2.1,0.9) rectangle (2.4, 1.2);
   \draw [draw=black, fill=yellow] (2.1,1.5) rectangle (2.4, 1.8);
   \draw [draw=black, fill=green] (2.1,2.1) rectangle (2.4, 2.4); 
   \draw [draw=black, fill=red] (2.1,2.7) rectangle (2.4, 3);
   \draw [draw=black, fill=yellow] (2.1,3.3) rectangle (2.4, 3.6);
   \draw [draw=black, fill=green] (2.1,0.3) rectangle (2.4, 0.6);

   \draw [draw=black, fill=red] (2.4,0.3) rectangle (2.7, 0.6); 
   \draw [draw=black, fill=yellow] (2.4,0.9) rectangle (2.7, 1.2);
   \draw [draw=black, fill=green] (2.4,1.5) rectangle (2.7, 1.8); 
   \draw [draw=black, fill=red] (2.4,2.1) rectangle (2.7, 2.4);
   \draw [draw=black, fill=yellow] (2.4,2.7) rectangle (2.7, 3);
   \draw [draw=black, fill=green] (2.4,3.3) rectangle (2.7, 3.6);
    
   \draw [draw=black, fill=red] (2.7,1.2) rectangle (3, 1.5);
   \draw [draw=black, fill=yellow] (2.7,1.8) rectangle (3, 2.1);
   \draw [draw=black, fill=green] (2.7,2.4) rectangle (3, 2.7); 
   \draw [draw=black, fill=red] (2.7,3) rectangle (3, 3.3);
   \draw [draw=black, fill=yellow] (2.7,0) rectangle (3, 0.3);
   \draw [draw=black, fill=green] (2.7,0.6) rectangle (3, 0.9);

  \draw [draw=black, fill=red] (3,2.4) rectangle (3.3, 2.7);
   \draw [draw=black, fill=yellow] (3,3) rectangle (3.3, 3.3);
   \draw [draw=black, fill=green] (3,0) rectangle (3.3, 0.3); 
   \draw [draw=black, fill=red] (3,0.6) rectangle (3.3, 0.9);
   \draw [draw=black, fill=yellow] (3,1.2) rectangle (3.3, 1.5);
   \draw [draw=black, fill=green] (3,1.8) rectangle (3.3, 2.1);

   \draw [draw=black, fill=red] (3.3,1.5) rectangle (3.6, 1.8);
   \draw [draw=black, fill=yellow] (3.3,2.1) rectangle (3.6, 2.4);
   \draw [draw=black, fill=green] (3.3,2.7) rectangle (3.6, 3); 
   \draw [draw=black, fill=red] (3.3,3.3) rectangle (3.6, 3.6);
   \draw [draw=black, fill=yellow] (3.3,0.3) rectangle (3.6, 0.6);
   \draw [draw=black, fill=green] (3.3,0.9) rectangle (3.6, 1.2);

 \node at (0.8, 4.15) { \small{TRP1} };
 \draw [draw=black, fill=red] (0.1,4) rectangle (0.4, 4.3);

 \node at (1.95, 4.15) { \small{TRP2} };
 \draw [draw=black, fill=yellow] (1.25,4) rectangle (1.55, 4.3);

  \node at (3.1, 4.15) { \small{TRP3} };
 \draw [draw=black, fill=green] (2.4,4) rectangle (2.7, 4.3);

  \node at (2, 3.8) { \scriptsize{Comb-6 DL-PRS} };

  \end{tikzpicture}
\end{subfigure}\hspace{0.2in}
      \begin{subfigure}[t]{0.45\linewidth}
      \begin{tikzpicture}

    \draw [->] (0,0) -- (4.35,0);
    \draw (0,0.3) -- (4.2,0.3);
    \draw (0,0.6) -- (4.2,0.6);
    \draw (0,0.9) -- (4.2,0.9);
    \draw (0,1.2) -- (4.2,1.2);
    \draw (0,1.5) -- (4.2,1.5);
    \draw (0,1.8) -- (4.2,1.8);
    \draw (0,2.1) -- (4.2,2.1);
    \draw (0,2.4) -- (4.2,2.4);
    \draw (0,2.7) -- (4.2,2.7);
    \draw (0,3) -- (4.2,3);
    \draw (0,3.3) -- (4.2,3.3);
    \draw (0,3.6) -- (4.2,3.6);

    \draw [->] (0,0) -- (0,3.8);
    \draw (0.3,0) -- (0.3, 3.6);
    \draw (0.6, 0) -- (0.6,3.6);
    \draw (0.9,0) -- (0.9,3.6);
    \draw (1.2,0) -- (1.2,3.6);
    \draw (1.5,0) -- (1.5,3.6);
    \draw (1.8,0) -- (1.8,3.6);
    \draw (2.1,0) -- (2.1,3.6);
    \draw (2.4,0) -- (2.4,3.6);
    \draw (2.7,0) -- (2.7,3.6);
    \draw (3.0,0) -- (3.0,3.6);
    \draw (3.3,0) -- (3.3,3.6);
    \draw (3.6,0) -- (3.6,3.6);
    \draw (3.9,0) -- (3.9,3.6);
    \draw (4.2,0) -- (4.2,3.6);
     
   \draw [draw=black, fill=blue] (0,0) rectangle (0.3, 0.3);
   \draw [draw=black, fill=blue] (0,1.2) rectangle (0.3, 1.5);
   \draw [draw=black, fill=blue] (0,2.4) rectangle (0.3, 2.7); 

   \draw [draw=black, fill=blue] (0.3,0.6) rectangle (0.6, 0.9);
   \draw [draw=black, fill=blue] (0.3,1.8) rectangle (0.6, 2.1);
   \draw [draw=black, fill=blue] (0.3,3) rectangle (0.6, 3.3); 
   
   \draw [draw=black, fill=blue] (0.6,0.3) rectangle (0.9, 0.6);
   \draw [draw=black, fill=blue] (0.6,1.5) rectangle (0.9, 1.8);
   \draw [draw=black, fill=blue] (0.6,2.7) rectangle (0.9, 3); 

   \draw [draw=black, fill=blue] (0.9,0.9) rectangle (1.2, 1.2);
   \draw [draw=black, fill=blue] (0.9,2.1) rectangle (1.2, 2.4);
   \draw [draw=black, fill=blue] (0.9,3.3) rectangle (1.2, 3.6);

   \draw [draw=black, fill=blue] (1.2,0) rectangle (1.5, 0.3);
   \draw [draw=black, fill=blue] (1.2,1.2) rectangle (1.5, 1.5);
   \draw [draw=black, fill=blue] (1.2,2.4) rectangle (1.5, 2.7); 

   \draw [draw=black, fill=blue] (1.5,0.6) rectangle (1.8, 0.9);
   \draw [draw=black, fill=blue] (1.5,1.8) rectangle (1.8, 2.1);
   \draw [draw=black, fill=blue] (1.5,3) rectangle (1.8, 3.3); 
   
   \draw [draw=black, fill=blue] (1.8,0.3) rectangle (2.1, 0.6);
   \draw [draw=black, fill=blue] (1.8,1.5) rectangle (2.1, 1.8);
   \draw [draw=black, fill=blue] (1.8,2.7) rectangle (2.1, 3); 

   \draw [draw=black, fill=blue] (2.1,0.9) rectangle (2.4, 1.2);
   \draw [draw=black, fill=blue] (2.1,2.1) rectangle (2.4, 2.4);
   \draw [draw=black, fill=blue] (2.1,3.3) rectangle (2.4, 3.6); 

   \draw [draw=black, fill=blue] (2.4,0) rectangle (2.7, 0.3);
   \draw [draw=black, fill=blue] (2.4,1.2) rectangle (2.7, 1.5);
   \draw [draw=black, fill=blue] (2.4,2.4) rectangle (2.7, 2.7); 

   \draw [draw=black, fill=blue] (2.7,0.6) rectangle (3, 0.9);
   \draw [draw=black, fill=blue] (2.7,1.8) rectangle (3, 2.1);
   \draw [draw=black, fill=blue] (2.7,3) rectangle (3, 3.3); 
   
   \draw [draw=black, fill=blue] (3,0.3) rectangle (3.3, 0.6);
   \draw [draw=black, fill=blue] (3,1.5) rectangle (3.3, 1.8);
   \draw [draw=black, fill=blue] (3,2.7) rectangle (3.3, 3); 

   \draw [draw=black, fill=blue] (3.3,0.9) rectangle (3.6, 1.2);
   \draw [draw=black, fill=blue] (3.3,2.1) rectangle (3.6, 2.4);
   \draw [draw=black, fill=blue] (3.3,3.3) rectangle (3.6, 3.6); 

     \node at (2, 3.8) { \scriptsize{Comb-4 UL-PRS} };

  \end{tikzpicture}
\end{subfigure}

\caption{Specific configurations of NR positioning reference signals. The DL-PRS is arranged in a comb-6 pattern with three TRPs. The UL-PRS from a UE has comb-4 pattern.}\label{PRS_SRS}
\end{figure}
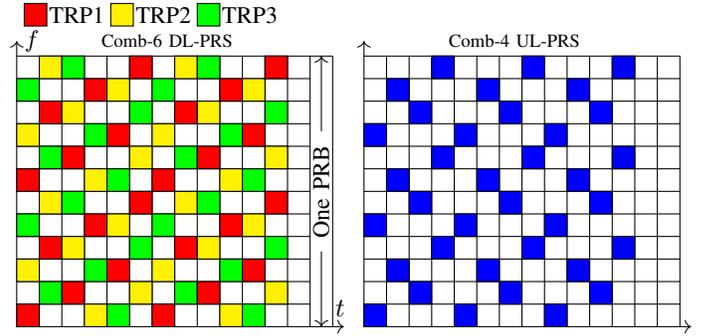

Multiple reference signals are used for communication related procedures. For the purpose of positioning, NR supports two new reference signals, the DL PRS and the UL SRS for positioning. Figure \ref{PRS_SRS} shows example within slot configurations of DL PRS and UL SRS for positioning.  The CSI-RS and SSB signals used for radio resource management (RRM) can also be used as part of the enhanced cell ID (E-CID) positioning method.

 \subsection{Downlink Positioning Reference Signal, DL-PRS} A NR DL-PRS can be configured at two levels, within a slot and at multi slot level. Within a slot, the starting resource element in time and frequency from a TRP can be configured. Across multiple slots, gaps between PRS slots, their periodicity and density within a period can be configured.        
  Here we explain some of the salient features of the DL-PRS specified in Release 16 \cite{38.211}. 
  \begin{enumerate}[wide, labelwidth=!, labelindent=0pt]

  \item \textbf{ Maximum Bandwidth:} The PRS footprint on the time frequency grid is configurable with a starting physical resource block (PRB) and a PRS bandwidth. The PRS may start at any PRB in the system bandwidth and can be configured with a bandwidth ranging from $24$ to $276$ PRBs in steps of $4$ PRBs. This amounts to a maximum bandwidth of about $100$\,MHz for $30$\,kHz subcarrier spacing and to about $400$\,MHz for $120$\,kHz subcarrier spacing. The flexible bandwidth configuration allows the network to configure the PRS while keeping out of band emissions to an acceptable level.
The large bandwidth allows a very significant improvement in time-of-arrival (TOA) accuracy compared to LTE.

\item \textbf{Resources and resource sets:} The PRS can be transmitted in beams. A PRS beam is referred to as a PRS resource while the full set of PRS beams transmitted from a TRP on the same frequency is referred to as a PRS resource set as illustrated in Fig.\,\ref{beams}.  The different beams can be time-multiplexed across symbols or slots. To assist UE RX beamforming, the DL PRS can be configured to be quasi-co-located (QCL) Type D with a DL reference signal from a serving or neighboring cell, signaling that the same RX beam used by the UE to receive said reference signal can be used to received the configured PRS.
The beam structure of the PRS improves coverage especially for mm-wave deployments and also allows for AoD estimation, e.g. the UE may measure DL PRS Received Signal Time Difference (RSTD) per beam and report the measured RSTD including DL PRS Resource id (beam id)  to the LMF. 

\begin{figure}[t]
  \begin{center}
    \includegraphics[width=0.5\textwidth]{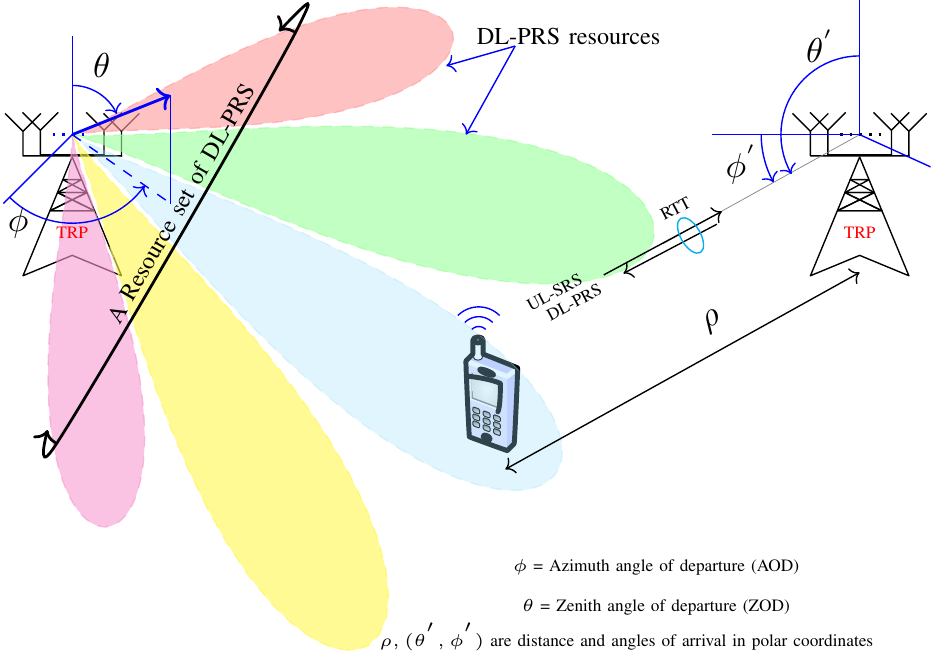}
  \end{center}
  \caption{An illustration showing a few positioning elements with NR. Beams as resources and set of beams as resource sets are shown. The newly supported positioning methods, multi-RTT and angle-based positioning methods are illustrated.    }\label{beams}
\end{figure}

 \item \textbf{Repetition and Periodicity:}      
In order to improve positioning accuracy, more measurements can be collected. Measurements are collected per resource. Hence, repeated transmission of PRS resources helps to collect more measurements. Measurements can be collected per resource.  The repetition of resources can be done in two ways, repeat before sweep and sweep before repeat as shown in the Fig.\ref{rep_mut}. The amount and type of repetition can be configured with parameters for configuring the gaps between resources ($T_{\text gap}^{\text PRS}$) and the number of resource repetition ($T_{\text rep}^{\text PRS}$) within a period of resource set ($T_{\text per}^{\text PRS}$).  The DL PRS resources can be repeated up to $32$ times within a resource set period, either in consecutive slots or with a configurable gap between repetitions. The resource set period in FR1 ranges from $4$ to $10240$ milli-seconds \cite{38.211}. For example, dense set of measurements can be collected by settings $T_{\text gap}^{\text PRS} = 1$,  $T_{\text rep}^{\text PRS} = 32$ and $T_{\text per}^{\text PRS} = 4$.

\item \textbf{Interference suppression:}  The DL PRS is designed to allow the UE to perform accurate TOA measurements in presence of interfering DL PRSs from nearby TRPs. Each symbol of the DL PRS has a comb-structure in frequency, i.e. the PRS utilizes every $N$th subcarrier. The comb value $N$ can be configured to be $2$, $4$, $6$ or $12$.  The length of the PRS within one slot is a multiple of $N$ symbols and the position of the first symbol within a slot is flexible as long as the slot consists of at-least $N$ PRS symbols. It allows accumulation of contiguous sub-carriers across a slot which improves correlation properties for TOA estimation.  The resource element pattern can be shifted in frequency with a frequency offset of $0$ to $N-1$ subcarriers thus allowing $N$ orthogonal DL PRSs utilizing the same symbols.  All configurable patterns cover every subcarriers in the configured bandwidth over the pattern duration which give maximum measurement range for the TOA measurement in scenarios with large delay spreads. The DL-PRS is QPSK modulated by a standardized 31-bit Gold code sequence initialized based on a DL PRS sequence ID taking values from $0$ to $4095$. As an example, the Fig.\ref{PRS_SRS} shows comb-6 DL-PRS and the pattern repeats after $6$ symbols. Figure \ref{PRS_SRS} shows multiplexing of three base stations to avoid interference among them.      The NR PRS comb-12 configuration allows for twice as many orthogonal signals as the comb-6 LTE PRS which is useful to mitigate interference. Further, the length of the NR PRS can be flexibly configured down to 2 symbols which can be useful e.g. in indoor scenarios where coverage is not an issue.

     Besides a comb structure allowing multiplexing of multiple TRPs in a slot, muting of signals can also be used as a way to mitigate interference. As shown in Fig.\ref{rep_mut}, muting can be used either at the repetition level, where each repetition can be individually muted within a periodic occasion, or at the occasion level, where the  whole periodic DL PRS occasion (including all repetitions) can be muted.

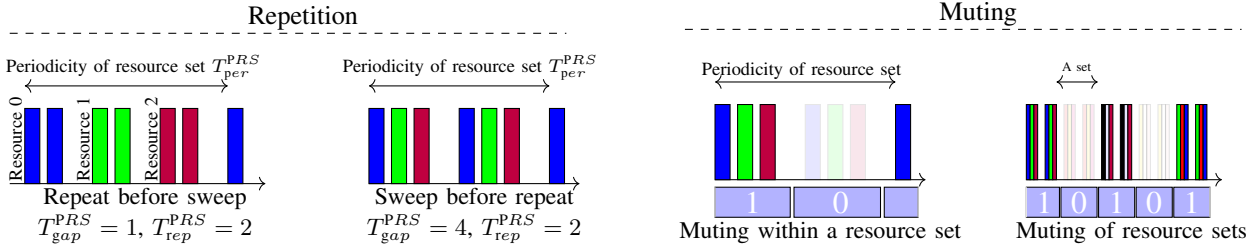
\begin{figure*}[t]
    \begin{center}
  \begin{subfigure}[t]{0.24\linewidth}
      \begin{tikzpicture}
        \draw [->] (-0.2,0) -- (3.2,0);
    \draw [draw=black, fill=blue] (0,0) rectangle (0.2, 1);
    \draw [draw=black, fill=blue] (0.3,0) rectangle (0.5, 1);
    \draw [draw=black, fill=green] (0.9,0) rectangle (1.1, 1);
    \draw [draw=black, fill=green] (1.2,0) rectangle (1.4, 1);
    \draw [draw=black, fill= purple] (1.8,0) rectangle (2.0, 1);
    \draw [draw=black, fill= purple] (2.1,0) rectangle (2.3, 1);
    \draw [draw=black, fill= blue] (2.7,0) rectangle (2.9, 1);
    \node at (-0.12, 0.6)[rotate=90] { \scriptsize{Resource 0} };
    \node at (0.78, 0.6)[rotate=90] { \scriptsize{Resource 1} };
    \node at (1.68, 0.6)[rotate=90] { \scriptsize{Resource 2} };
        \draw [<->] (0,1.3) -- (2.7,1.3);
        \node at (1.45, 1.55){ \scriptsize{Periodicity of resource set $T_{\text per}^{\text PRS}$} };
        
        \node at (1.6, -0.2){ \small{Repeat before sweep}};
        \node at (1.6, -0.6){ \small{$T_{\text gap}^{\text PRS} = 1$, $T_{\text rep}^{\text PRS} = 2$ }};
        \draw [-, dashed] (-0.2,2) -- (7.2,2);
\node at (3.7, 2.2){ Repetition};
      \end{tikzpicture}
    \end{subfigure}
  \begin{subfigure}[t]{0.24\linewidth}
      \begin{tikzpicture}
    \draw [->] (0.2,0) -- (3,0);
    \draw [draw=black, fill=blue] (0.2,0) rectangle (0.4, 1);
    \draw [draw=black, fill= green] (0.5,0) rectangle (0.7, 1);
    \draw [draw=black, fill=purple] (0.8,0) rectangle (1.0, 1);
    \draw [draw=black, fill=blue] (1.4,0) rectangle (1.6, 1);
    \draw [draw=black, fill= green] (1.7,0) rectangle (1.9, 1);
    \draw [draw=black, fill= purple] (2.0,0) rectangle (2.2, 1);
    \draw [draw=black, fill= blue] (2.6,0) rectangle (2.8, 1);
    \draw [<->] (0.2,1.3) -- (2.6,1.3);
    \node at (1.55, 1.55){ \scriptsize{Periodicity of resource set $T_{\text per}^{\text PRS}$} };
        \node at (1.6, -0.2){ \small{Sweep before repeat }};
        \node at (1.6, -0.6){ \small{$T_{\text gap}^{\text PRS} = 4$, $T_{\text rep}^{\text PRS} = 2$}};

  \end{tikzpicture}
  \end{subfigure}
  \begin{subfigure}[t]{0.24\linewidth}
    \begin{tikzpicture}
    \draw [->] (0.2,0) -- (3,0);
    \draw [draw=black, fill=blue] (0.2,0) rectangle (0.4, 1);
    \draw [draw=black, fill= green] (0.5,0) rectangle (0.7, 1);
    \draw [draw=black, fill=purple] (0.8,0) rectangle (1.0, 1);
    \draw [draw=black, fill=blue, opacity=0.1] (1.4,0) rectangle (1.6, 1);
    \draw [draw=black, fill= green, opacity=0.1] (1.7,0) rectangle (1.9, 1);
    \draw [draw=black, fill= purple, opacity=0.1] (2.0,0) rectangle (2.2, 1);
    \draw [draw=black, fill= blue] (2.6,0) rectangle (2.8, 1);
    \draw [<->] (0.2,1.3) -- (2.6,1.3);
    \node at (1.35, 1.45){ \scriptsize{Periodicity of resource set} };
    \draw [draw=black, fill=blue!30] (0.2,-0.5) rectangle (1.2, -0.1);
    \draw [draw=black, fill=blue!30] (1.25,-0.5) rectangle (2.4, -0.1);
    \draw [draw=black, fill=blue!30] (2.45,-0.5) rectangle (2.9, -0.1);
    \draw [-][color=white] (0.2,-0.5) -- (0.2,-0.1);
    \draw [-][color=white] (2.9,-0.5) -- (2.9,-0.1);
    \node [color=white] at (0.7, -0.3){ \large 1 };
    \node [color=white] at (1.87, -0.3){ \large 0 };
    \node at (1.6, -0.7){ \small{Muting within a resource set}};

        \draw [-, dashed] (-0.2,2) -- (7.2,2);
\node at (3.7, 2.2){ Muting};

  \end{tikzpicture}
    \end{subfigure}
  \begin{subfigure}[t]{0.24\linewidth}
      \begin{tikzpicture}
    \draw [->] (0.2,0) -- (2.8,0);
    \draw [draw=black, fill=blue] (0.2,0) rectangle (0.25, 1);
    \draw [draw=black, fill= green] (0.25,0) rectangle (0.3, 1);
    \draw [draw=black, fill=purple] (0.3,0) rectangle (0.35, 1);

    \draw [draw=black, fill=blue] (0.45,0) rectangle (0.5, 1);
    \draw [draw=black, fill= green] (0.5,0) rectangle (0.55, 1);
    \draw [draw=black, fill=purple] (0.55,0) rectangle (0.6, 1);

    \draw [draw=black, fill=red, opacity=0.1] (0.7,0) rectangle (0.75, 1);
    \draw [draw=black, fill= yellow, opacity=0.1] (0.75,0) rectangle (0.8, 1);
    \draw [draw=black, fill=magenta,opacity=0.1] (0.8,0) rectangle (0.85, 1);

    \draw [draw=black, fill=red,opacity=0.1] (0.95,0) rectangle (1, 1);
    \draw [draw=black, fill= yellow,opacity=0.1] (1,0) rectangle (1.05, 1);
    \draw [draw=black, fill=magenta,opacity=0.1] (1.05,0) rectangle (1.1, 1);

    \draw [draw=black, fill=black] (1.2,0) rectangle (1.25, 1);
    \draw [draw=black, fill= blue!10] (1.25,0) rectangle (1.3, 1);
    \draw [draw=black, fill=purple] (1.3,0) rectangle (1.35, 1);

    \draw [draw=black, fill=black] (1.45,0) rectangle (1.5, 1);
    \draw [draw=black, fill= blue!10] (1.5,0) rectangle (1.55, 1);
    \draw [draw=black, fill=purple] (1.55,0) rectangle (1.6, 1);

    \draw [draw=black, fill=yellow,opacity=0.1] (1.7,0) rectangle (1.75, 1);
    \draw [draw=black, fill= white,opacity=0.1] (1.75,0) rectangle (1.8, 1);
    \draw [draw=black, fill=red!20,opacity=0.1] (1.8,0) rectangle (1.85, 1);
    
    \draw [draw=black, fill=yellow,opacity=0.1] (1.95,0) rectangle (2, 1);
    \draw [draw=black, fill= white,opacity=0.1] (2.0,0) rectangle (2.05, 1);
    \draw [draw=black, fill=red!20,opacity=0.1] (2.05,0) rectangle (2.1, 1);

    \draw [draw=black, fill=green] (2.2,0) rectangle (2.25, 1);
    \draw [draw=black, fill= red] (2.25,0) rectangle (2.3, 1);
    \draw [draw=black, fill=blue] (2.3,0) rectangle (2.35, 1);

    \draw [draw=black, fill=green] (2.45,0) rectangle (2.5, 1);
    \draw [draw=black, fill= red] (2.5,0) rectangle (2.55, 1);
    \draw [draw=black, fill=blue] (2.55,0) rectangle (2.6, 1);
    
    \draw [<->] (0.6,1.3) -- (1.15,1.3);
    \node at (0.87, 1.5){ \tiny{A set} };
    \draw [draw=black, fill=blue!30] (0.2,-0.5) rectangle (0.64, -0.1);
    \draw [draw=black, fill=blue!30] (0.66,-0.5) rectangle (1.14, -0.1);
    \draw [draw=black, fill=blue!30] (1.16,-0.5) rectangle (1.64, -0.1);
    \draw [draw=black, fill=blue!30] (1.66,-0.5) rectangle (2.14, -0.1);
    \draw [draw=black, fill=blue!30] (2.16,-0.5) rectangle (2.64, -0.1);

    \draw [-][color=white] (0.2,-0.5) -- (0.2,-0.1);
    \draw [-][color=white] (2.64,-0.5) -- (2.64,-0.1);
    \node [color=white] at (0.44, -0.3){ \large 1 };
    \node [color=white] at (0.9, -0.3){ \large 0 };
    \node [color=white] at (1.4, -0.3){ \large 1 };
   \node [color=white] at (1.9, -0.3){ \large 0 };
   \node [color=white] at (2.4, -0.3){ \large 1 };
         \node at (1.6, -0.7){ \small{Muting of resource sets}};      
  \end{tikzpicture}
    \end{subfigure}
  \end{center}
 \caption{Repetition and muting of PRS resources and resource sets. PRS resources and resource sets can be repeated for improving accuracy by enabling collection of more measurements. However, they can be muted too for reducing interference.  }\label{rep_mut}
\end{figure*}

 \item \textbf{Hierarchical structure:} The DL-PRS configuration is provided in a hierarchy as shown in Fig.\ref{hierarchy}. There can be at most $4$ frequency layers and each frequency layer has at most $64$ TRPs. Each TRP per frequency layer can have $2$ DL-PRS Resource sets thus resulting in a total of $8$ resource sets per TRP and each resource set can have up to $64$ resources. Each resource corresponds to a beam. Having $2$ different resource sets per frequency layer per TRP allows gNB to configure one set of wide beams and another set of narrow beams for each frequency layer.  

      \end{enumerate}

\subsection{Uplink signal, UL-SRS for positioning}

The SRS for positioning is a reference signal based on the SRS for communication. Although the signals have a lot in common, SRS for positioning and for communication are configured separately and with different properties specific to their usage. The UL-SRS shown in the Fig.\ref{PRS_SRS}, is a comb-4 signal. 
 \begin{enumerate}[wide, labelwidth=!, labelindent=0pt]
  \item \textbf{Resources and resource sets:} The SRS for positioning is configured in a resource, which in turn can be part of a resource set. a resource correspond to an SRS beam, and resource sets correspond to a collection of SRS resource (i.e. beams)  aimed at a  given TRP. 
  
  \item \textbf{Coverage features}   The SRS resource is defined as a collection of symbols transmitted on the time-frequency NR grid. Like the DL PRS, the SRS resources for positioning  are transmitted on a single antenna port, and can be placed to begin on any symbol in the NR uplink slot. In the time domain, the SRS resources for positioning can span {1,2,4,8,12} consecutive OFDM symbols which provide enough coverage to reach all TRPs involved in the positioning procedures.  Contrary to the SRS for communication, repetition is not supported in an SRS for positioning resource. Similar to SRS for communication, the SRS for positioning is using Zadoff-Chu sequences as a base signal, to ensure low-PAPR transmission from the UE. The particular sequence used to generate an SRS symbol depends on configuration parameters and sequence hopping is supported as for the SRS for communication. 

  \item \textbf{Interference-free UE multiplexing} Since 3GPP Release 15, NR has used a comb structure for the SRS, so that only a fraction of the OFDM subcarriers are occupied by a given SRS resource. For the SRS for communication, the comb size is either 2 or 4, meaning that the SRS occupies 1 subcarrier out of 2 or 4, respectively. In Release 16, several enhancements were added in the specification of the SRS for positioning. The comb size $K_{TC}$  for the SRS for positioning is {2,4,8} and new comb patterns was specified in order for all the subcarriers to be sounded in one resource. Different resource element offsets pattern can be configured as a function of the comb size $K_{TC}$ and the number of symbol in the resource $N_{\text{symb}}^{\text{SRS}}$ \cite{38.211}. For UE multiplexing, SRS for positioning can be configured with an initial comb offset and a specific cyclic shift. The comb offset can take any value in. $[0, K_{TC} - 1$]. 

  \end{enumerate}

\begin{figure*}[!t]
  \begin{center}
  \begin{tikzpicture}
    \draw [fill=blue!30](0,3)circle(0.4cm);
    \node [color=black] at (0, 3){\scriptsize $\text{ \scriptsize PFL}_1$ };

    \draw [fill=green!30](-1.3,2)circle(0.4cm);
    \node [color=black] at (-1.3, 2.1){\scriptsize $\text{ \scriptsize TRP}_1$ };
    \node [color=black] at (-1.3, 1.8){\tiny $ \text{\tiny ID}_1$ };
    
    \draw [fill=green!30](-0.4,2)circle(0.4cm);
    \node [color=black] at (-0.4, 2.1){\scriptsize $\text{ \scriptsize TRP}_2$  };
    \node [color=black] at (-0.4, 1.8){\tiny $ \text{\tiny ID}_2$ };

    \draw [fill=green!30](1,2)circle(0.42cm);
    \node [color=black] at (1, 2.1){\scriptsize $\text{\scriptsize TRP}_{64}$  };
    \node [color=black] at (1, 1.8){\tiny $ \text{\tiny ID}_{64}$ };

    \draw [->] (-0.32,2.8) -- (-1.1,2.35);
    \draw [->] (-0.17,2.65) -- (-0.35,2.4);
    \draw [->] (0.3,2.75) -- (0.75,2.3);
   \draw [dotted, thick] (0.1,2) -- (0.55,2);

    \draw [fill=red!30](-1.7,1)circle(0.35cm);
    \node [color=black] at (-1.7,1){\tiny $\text{\tiny Set ID}_{1}$  };

    \draw [fill=gray!30](-2.7,0.1)circle(0.35cm);
    \node [color=black] at (-2.7,0.1){\tiny $\text{\tiny R ID}_{1}$  };

    \draw [fill=gray!30](-1.95,0.1)circle(0.35cm);
    \node [color=black] at (-1.95,0.1){\tiny $\text{\tiny R ID}_{2}$  };
    \draw [fill=gray!30](-0.8,0.1)circle(0.35cm);
    \node [color=black] at (-0.8,0.1){\tiny $\text{\tiny R ID}_{64}$  };
    
    \draw [->] (-2.0,0.8) -- (-2.5,0.4);
    \draw [->] (-1.75,0.68) -- (-1.95,0.5);
    \draw [->] (-1.6,0.7) -- (-1.1,0.25);

   \draw [dotted, thick] (-1.5,0.1) -- (-1.25,0.1);

      \draw [dotted, thick] (0.2,0.1) -- (1.2,0.1);

    \draw [fill=red!30](-0.9,1)circle(0.35cm);
    \node [color=black] at (-0.9,1){\tiny $\text{\tiny Set ID}_{2}$  };

    \draw [->] (-1.55,1.7) -- (-1.65,1.35);
    \draw [->] (-1.05,1.7) -- (-0.95,1.35);

      \draw [dotted, thick] (-0.4,1) -- (0.1,1);

    \draw [fill=red!30](0.6,1)circle(0.35cm);
    \node [color=black] at (0.6,1){\tiny $\text{\tiny Set ID}_{1}$  };

    \draw [fill=red!30](1.4,1)circle(0.35cm);
    \node [color=black] at (1.4,1){\tiny $\text{\tiny Set ID}_{2}$  };

    \draw [->] (0.75,1.7) -- (0.65,1.35);
    \draw [->] (1.25,1.7) -- (1.35,1.35);

        \draw [fill=blue!30](4,3)circle(0.4cm);
    \node [color=black] at (4, 3){\scriptsize $\text{ \scriptsize PFL}_2$ };

    \draw [fill=green!30](2.7,2)circle(0.4cm);
     \node [color=black] at (2.7, 2.1){\scriptsize $\text{ \scriptsize TRP}_{1}$  };
    \node [color=black] at (2.7, 1.8){\tiny $ \text{\tiny ID}_{65}$ };

    \draw [fill=green!30](3.6,2)circle(0.4cm);
      \node [color=black] at (3.6, 2.1){\scriptsize $\text{ \scriptsize TRP}_{2}$  };
    \node [color=black] at (3.6, 1.8){\tiny $ \text{\tiny ID}_{66}$ };

   \draw [dotted, thick] (2,1) -- (2.6,1);

    \draw [fill=red!30](3.2,1)circle(0.35cm);
    \node [color=black] at (3.2,1){\tiny $\text{\tiny Set ID}_{3}$  };


    \draw [fill=gray!30](2.2,0.1)circle(0.35cm);
    \node [color=black] at (2.2,0.1){\tiny $\text{\tiny R ID}_{1}$  };

    \draw [fill=gray!30](2.95,0.1)circle(0.35cm);
    \node [color=black] at (2.95,0.1){\tiny $\text{\tiny R ID}_{2}$  };
    \draw [fill=gray!30](4.1,0.1)circle(0.35cm);
    \node [color=black] at (4.1,0.1){\tiny $\text{\tiny R ID}_{64}$  };
    
    \draw [->] (2.9,0.8) -- (2.4,0.4);
    \draw [->] (3.25,0.68) -- (2.95,0.5);
    \draw [->] (3.3,0.7) -- (3.8,0.25);
    
   \draw [dotted, thick] (3.4,0.1) -- (3.65,0.1);

   \draw [dotted, thick] (5,0.1) -- (6,0.1);

    \draw [fill=red!30](4,1)circle(0.35cm);
    \node [color=black] at (4,1){\tiny $\text{\tiny Set ID}_{4}$  };

    \draw [->] (3.35,1.7) -- (3.25,1.35);
    \draw [->] (3.85,1.7) -- (3.95,1.35);


    \draw [fill=green!30](5,2)circle(0.4cm);
    \node [color=black] at (5, 2.1){\scriptsize $\text{ \scriptsize TRP}_{64}$  };
    \node [color=black] at (5, 1.8){\tiny $ \text{\tiny ID}_{128}$ };

    \draw [->] (3.68,2.8) -- (2.9,2.35);
    \draw [->] (3.83,2.65) -- (3.65,2.4);
    \draw [->] (4.3,2.75) -- (4.75,2.3);
   \draw [dotted, thick] (4.1,2) -- (4.55,2);

    \draw [fill=blue!30](8,3)circle(0.4cm);
    \node [color=black] at (8, 3){\scriptsize $\text{ \scriptsize PFL}_3$ };

    \draw [fill=green!30](6.7,2)circle(0.4cm);
    \node [color=black] at (6.7, 2.1){\scriptsize $\text{ \scriptsize TRP}_{1}$  };
    \node [color=black] at (6.7, 1.8){\tiny $ \text{\tiny ID}_{129}$ };
    
    \draw [fill=green!30](7.6,2)circle(0.4cm);
    \node [color=black] at (7.6, 2.1){\scriptsize $\text{ \scriptsize TRP}_{2}$  };
    \node [color=black] at (7.6, 1.8){\tiny $ \text{\tiny ID}_{130}$ };

   \draw [fill=red!30](7.2,1)circle(0.35cm);
    \node [color=black] at (7.2,1){\tiny $\text{\tiny Set ID}_{5}$  };

    \draw [fill=red!30](8,1)circle(0.35cm);
    \node [color=black] at (8,1){\tiny $\text{\tiny Set ID}_{6}$  };

   \draw [dotted, thick] (8.1,2) -- (8.5,2);


    \draw [fill=gray!30](7.0,0.1)circle(0.35cm);
    \node [color=black] at (7.0,0.1){\tiny $\text{\tiny R ID}_{1}$  };

    \draw [fill=gray!30](7.75,0.1)circle(0.35cm);
    \node [color=black] at (7.75,0.1){\tiny $\text{\tiny R ID}_{2}$  };
    \draw [fill=gray!30](8.9,0.1)circle(0.35cm);
    \node [color=black] at (8.9,0.1){\tiny $\text{\tiny R ID}_{64}$  };

   \draw [dotted, thick] (8.2,0.1) -- (8.45,0.1);
    
    \draw [->] (7.7,0.8) -- (7.2,0.4);
    \draw [->] (8.05,0.68) -- (7.75,0.5);
    \draw [->] (8.1,0.7) -- (8.6,0.25);

    \draw [->] (7.35,1.7) -- (7.25,1.35);
    \draw [->] (7.85,1.7) -- (7.95,1.35);

      \draw [dotted, thick] (5,1) -- (6.1,1);
      \draw [dotted, thick] (9,1) -- (11.5,1);
      \draw [dotted, thick] (10,0.1) -- (11,0.1);

    \draw [fill=green!30](9,2)circle(0.4cm);
    \node [color=black] at (9, 2.1){\scriptsize $\text{ \scriptsize TRP}_{64}$  };
    \node [color=black] at (9, 1.8){\tiny $ \text{\tiny ID}_{192}$ };

    \draw [->] (7.68,2.8) -- (6.9,2.35);
    \draw [->] (7.83,2.65) -- (7.65,2.4);
    \draw [->] (8.3,2.75) -- (8.75,2.3);

        \draw [fill=blue!30](12,3)circle(0.4cm);
    \node [color=black] at (12, 3){\scriptsize $\text{ \scriptsize PFL}_4$ };

    \draw [fill=green!30](10.7,2)circle(0.4cm);
    \node [color=black] at (10.7, 2.1){\scriptsize $\text{ \scriptsize TRP}_{1}$  };
    \node [color=black] at (10.7, 1.8){\tiny $ \text{\tiny ID}_{193}$ };
    
    \draw [fill=green!30](11.6,2)circle(0.4cm);
    \node [color=black] at (11.6, 2.1){\scriptsize $\text{ \scriptsize TRP}_{2}$  };
    \node [color=black] at (11.6, 1.8){\tiny $ \text{\tiny ID}_{194}$ };
    
    \draw [fill=green!30](13,2)circle(0.4cm);
    \node [color=black] at (13, 2.1){\scriptsize $\text{ \scriptsize TRP}_{64}$  };
    \node [color=black] at (13, 1.8){\tiny $ \text{\tiny ID}_{256}$ };

    \draw [->] (11.68,2.8) -- (10.9,2.35);
    \draw [->] (11.83,2.65) -- (11.65,2.4);
    \draw [->] (12.3,2.75) -- (12.75,2.3);
   \draw [dotted, thick] (12.1,2) -- (12.55,2);


    \draw [fill=red!30](12.6,1)circle(0.35cm);
    \node [color=black] at (12.6,1){\tiny $\text{\tiny Set ID}_{7}$  };

    \draw [fill=red!30](13.4,1)circle(0.35cm);
    \node [color=black] at (13.4,1){\tiny $\text{\tiny Set ID}_{8}$  };

    \draw [->] (12.75,1.7) -- (12.65,1.35);
    \draw [->] (13.25,1.7) -- (13.35,1.35);

    \draw [fill=gray!30](12.4,0.1)circle(0.35cm);
    \node [color=black] at (12.4,0.1){\tiny $\text{\tiny R ID}_{1}$  };

    \draw [fill=gray!30](13.15,0.1)circle(0.35cm);
    \node [color=black] at (13.15,0.1){\tiny $\text{\tiny R ID}_{2}$  };
    \draw [fill=gray!30](14.3,0.1)circle(0.35cm);
    \node [color=black] at (14.3,0.1){\tiny $\text{\tiny R ID}_{64}$  };

   \draw [dotted, thick] (13.6,0.1) -- (13.85,0.1);

    \draw [->] (13.1,0.8) -- (12.6,0.4);
    \draw [->] (13.45,0.68) -- (13.15,0.5);
    \draw [->] (13.5,0.7) -- (14,0.25);
    
  \end{tikzpicture}
\end{center}

\caption{NR DL-PRS configuration hierarchy. The configuration hierarchy allows network to provide assistance data in a structured format. It also enables the UE to locate the resources unambiguously to perform measurements.}\label{hierarchy}
  \end{figure*}
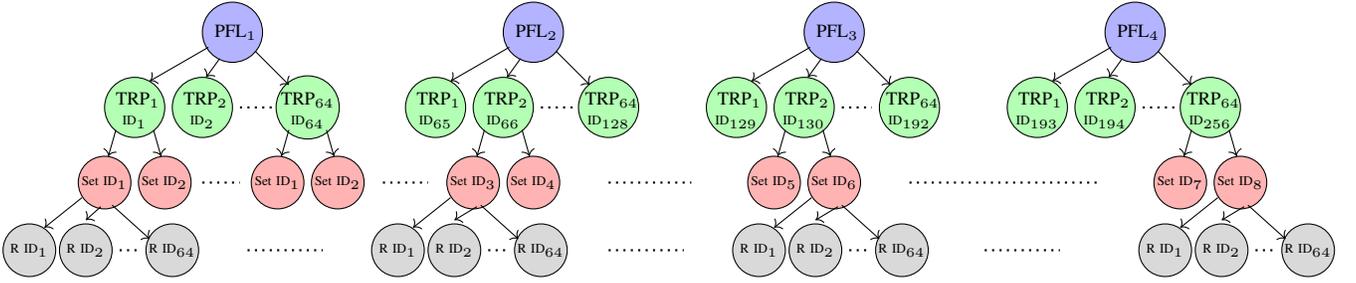

\section{Measurements for Positioning}

Compared to LTE, more positioning measurements have been standardized for NR. Furthermore, some of the measurement types (e.g., Rx-Tx time difference) have been standardized not only for the serving cell but also for neighbor cells or TRPs. The following downlink  measurements based on PRS can be configured in the UE by LMF: RSTD, UE Rx-Tx time difference, and PRS-based Reference Signal Received Power (PRS-RSRP) measurements, unlike in LTE where only RSTD were possible for PRS. In addition, the UE may also be requested to report some RRM measurements to support E-CID positioning method: Synchronization Signal based RSRP (SS-RSRP),  Synchronization  Signal  based  Reference Signal Received Quality (SS-RSRQ), Channel-State Information Reference Signal  based RSRP (CSI-RSRP), and Channel-State Information Reference Signal based RSRQ (CSI-RSRQ). The following UL measurements can be configured by LMF and reported by gNB: UL Relative Time of Arrival (UL-RTOA), gNB Rx-Tx time difference, Sounding Reference Signal RSRP (SRS-RSRP), Azimuth and Zenith of Angle of Arrival (A-AoA and Z-AoA, respectively). The 3GPP standard supports the above measurements in all supported bands and both lower and higher frequency ranges (FR1 and FR2, respectively), over the corresponding bandwidths within the operating frequency bands, which are up to $100$\,MHz in FR1 and $400$\,MHz in FR2.

The range of reportable absolute values for power-based measurements is $[-156; -31]$\,dBm, with $1$\,dB resolution, similar to that in LTE. The range of reportable absolute values for timing-based measurements is $[-985024; 985024]$ in $T_c$ units, with a flexible resolution step of $2kT_c$,where  $1 T_c$  corresponds to $0.51\,\text{ns}$ and $k$ is an integer in the interval $[2; 5]$ for FR1 and $[0; 5]$ for FR2. The value of $k$ can be configured by LMF or adjusted by the UE, e.g., to account for the NR subcarrier spacing and the UE reporting capability.

The above measurements can be performed for serving and neighbor TRPs and can be used for a variety of RAT-dependent and hybrid positioning methods, standardized or not. Some examples of the standardized methods are \cite{38.215},

\begin{itemize}
\item DL Time Difference of Arrival (DL-TDOA) - based on RSTD and optional complimentary PRS-RSRP;
\item DL Angle of Departure (DL-AoD) - based on PRS-RSRP;
\item UL-TDOA - based on UL-RTOA and optional complimentary SRS-RSRP;  
\item UL-AoA – based on A-AoA and Z-AoA and optional complimentary SRS-RSRP measurements; 
\item 	Multi-RTT – based on gNB Rx-Tx and UE Rx-Tx time difference measurements and optional complimentary PRS-RSRP, SRS-RSRP, A-AoA, and Z-AoA measurements.
\end{itemize}

Accurate and timely measurements are necessary to ensure reliable positioning, therefore there are standardized requirements on the maximum allowed time during which the measurements are to be performed (a.k.a. measurement period) and the maximum allowed error for the reported measurements (a.k.a. measurement accuracy). The measurement period spans the time necessary to obtain no more than four measurement samples, while the achieved accuracy should not be worse than the corresponding measurement accuracy requirement, which is expected to be more stringent than that in LTE.

\section{Simulations and Discussions}
Based on above discussions on various aspects of NR positioning as specified in Release 16, we demonstrate the performance of positioning in this section.
We have evaluated 5G positioning performance in three different scenarios using 3GPP channel models Urban Macro (UMa), Urban Micro (UMi) and Indoor Open Office (IOO) \cite{38.901}. The scenarios for evaluation is taken from agreed evaluation scenarios during the Release 16 positioning standardization. Figure\,\ref{fig:sim} shows simulation results for UMa and UMi scenarios with different dimensions.  The UMa and UMi scenarios have $1600m\times 1600m$ and  $500m\times 500m$ areas respectively. The inter-site distances(ISD) are $500$, $200$ and $20$ meters in UMa, UMi and IOO scenarios respectively. The UMa and UMi scenarios consists of $7$ hexagonal cells, with each cell consisting of $3$ sectors. Hence the number of TRPs in UMa and UMi scenarios are $21$. Whereas, in IOO it is $12$.  These scenarios are described in \cite{38.901, 38.855}.  The evaluations are done for UMa and UMi in FR1 and the IOO scenario is evaluated both in FR1 and FR2. The evaluations are done both in presence and absence of interferences. Positioning methods in UMa and UMi scenarios is DL-TDOA based positioning. Whereas, in IOO we have also shown performances using multi-RTT and uplink angle of arrival (UL-AOA) based methods. The network is assumed ideally time synchronized for TDOA based positioning evaluations. The simulations are done for maximum bandwidth corresponding to $272$ PRBs both in uplink and downlink.

\subsection{TRPs and the downlink simulation parameters}

The downlink transmitted powers are $49, 42$ and $23$\,dBm respectively in the three scenarios. The carrier frequency chosen in FR1 is $2$\,GHz and in FR2 it is $28$\,GHz. The subcarrier spacings for FR1 and FR2 is chosen to be $30$\,kHz and $120$\,kHz  respectively. The DL-PRS used in simulation has comb-12 configuration spanning across $12$ symbols of a slot.  The receiver noise figure for receiving uplink signals is assumed $5$\,dB. The TRP heights are $20-50$\,m uniformly distributed in UMa, $10$\,m in UMi and $3$\,m in IOO. The TRP antennas are explained in \cite{38.855}.

\subsection{The UE and the uplink  simulation parameters}

The UL-PRS for multi-RTT and UL-AoA methods have comb-2 configuration spanning across $2$ symbols of a slot. The receiver noise figure is assumed $9$\,dB. The UE speeds are assumed to be $60$\,km/h in UMa and  $3$\,km/h in UMi and IOO. The UE antenna is assumed dual-polarized isotropical.

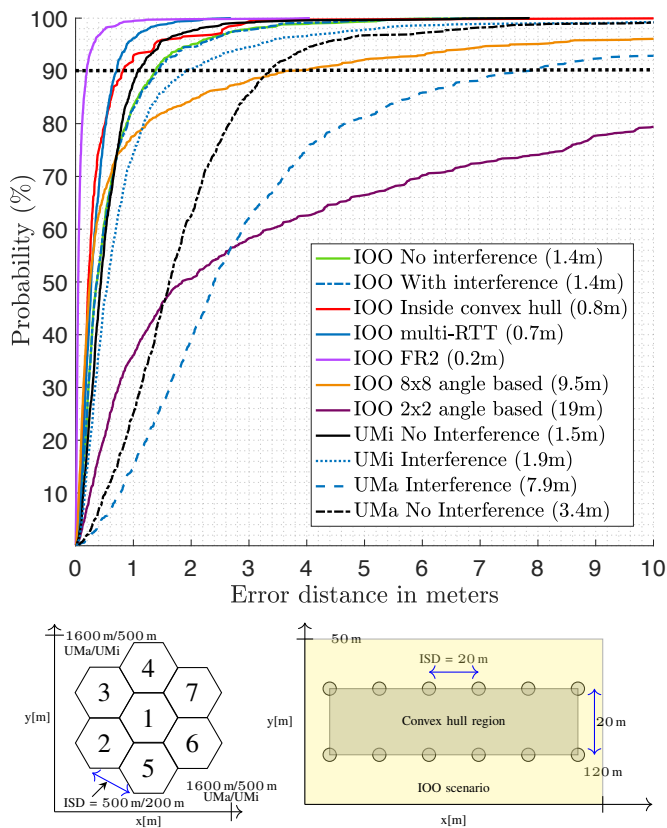
\begin{figure}[t!]
   \includegraphics[width=1\linewidth]{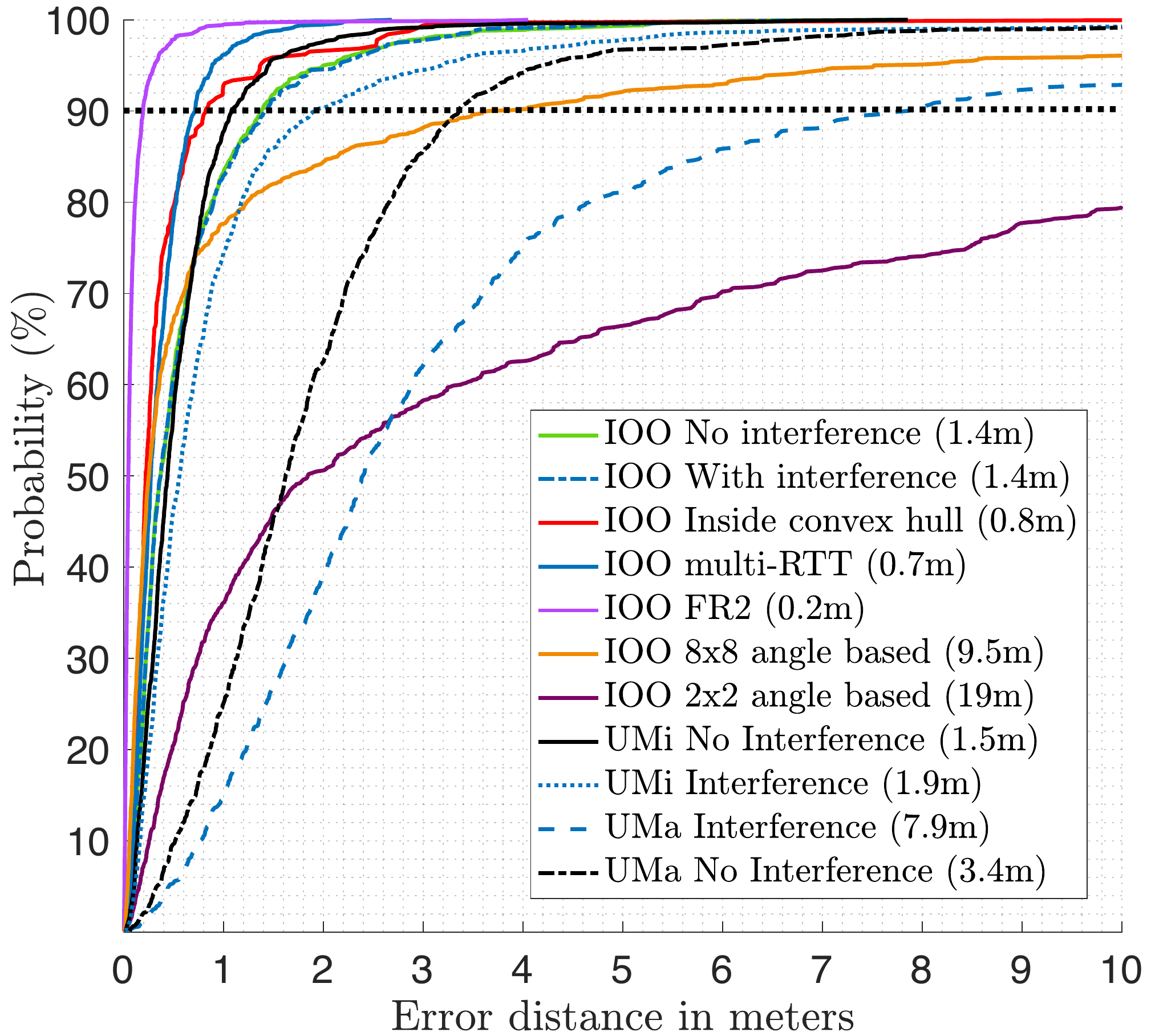}
    \begin{center}
  \begin{subfigure}{0.35\linewidth} 
\begin{tikzpicture}[scale=0.5]
  \node[regular polygon, regular polygon sides=6, shape aspect=1, minimum width=1, minimum height=1, draw] at (0,0){ 1};
   \node[regular polygon, regular polygon sides=6, shape aspect=1, minimum width=0.1cm, minimum height=1, draw]at (-1.17,-0.67) {2};
   \node[regular polygon, regular polygon sides=6, shape aspect=1, minimum width=0.1cm, minimum height=1, draw]at (-1.17,0.67) {3};
   \node[regular polygon, regular polygon sides=6, shape aspect=1, minimum width=0.1cm, minimum height=1, draw]at (0,1.33) {4};
   \node[regular polygon, regular polygon sides=6, shape aspect=1, minimum width=0.1cm, minimum height=1, draw]at (0,-1.33) {5};
   \node[regular polygon, regular polygon sides=6, shape aspect=1, minimum width=0.1cm, minimum height=1, draw]at (1.17,-0.67) {6};
   \node[regular polygon, regular polygon sides=6, shape aspect=1, minimum width=0.1cm, minimum height=1, draw]at (1.17, 0.67) {7};
         \draw [-] (-2.75, 2.2) -- (-2.25,2.2);      
         \draw [-] (2.2,-2.75) -- (2.2,-2.25);

   \draw [->] (-2.5,-2.5) -- (-2.5,2.5);      
   \draw [->] (-2.5,-2.5) -- (2.5,-2.5);

    \node [color=black] at (-1.0,2.2){\tiny $1600$\,m/$500$\,m};
    \node [color=black] at (-1.5,1.8){\tiny UMa/UMi};
    \node [color=black] at (2.2,-1.8){\tiny $1600$\,m/$500$\,m};
    \node [color=black] at (2.2,-2.1){\tiny UMa/UMi};
    \node [color=black] at (-2.95,0){\tiny y[m]  };
    \node [color=black] at (0,-2.75){\tiny x[m]  };

    \node [color=black] at (-0.65,-2.25){\tiny ISD = $500$\,m/$200$\,m };
   \draw [<->, color=blue] (-1.5,-1.46) -- (-0.5,-2);      
   \draw [->, >=stealth]  (-1.5,-2.1)-- (-1.1,-1.7);

\end{tikzpicture}
    \end{subfigure}
  \begin{subfigure}{0.6\linewidth}
      \begin{tikzpicture}[scale=2.2]
        \draw [->] (0,0) -- (0,1.1);      
        \draw [->] (0,0) -- (2.1,0);
         \draw [-] (1.8,-0.05) -- (1.8,0.05);      
         \draw [-] (-0.05,1) -- (0.05,1);
    \node [color=black] at (0.25,1){\tiny $50$\,m  };
    \node [color=black] at (1.8,0.2){\tiny $120$\,m  };
    \node [color=black] at (-0.15,0.5){\tiny y[m]  };
    \node [color=black] at (0.9,-0.1){\tiny x[m]  };
    \node [color=black] at (0.9,0.88){\tiny ISD = $20$\,m  };
       
    \draw [fill=gray!30](0.15,0.3)circle(0.04cm);
    \draw [fill=gray!30](0.45,0.3)circle(0.04cm);
    \draw [fill=gray!30](0.75,0.3)circle(0.04cm);
    \draw [fill=gray!30](1.05,0.3)circle(0.04cm);
    \draw [fill=gray!30](1.35,0.3)circle(0.04cm);
    \draw [fill=gray!30](1.65,0.3)circle(0.04cm);
    \draw [fill=gray!30](0.15,0.7)circle(0.04cm);
    \draw [fill=gray!30](0.45,0.7)circle(0.04cm);
    \draw [fill=gray!30](0.75,0.7)circle(0.04cm);
    \draw [fill=gray!30](1.05,0.7)circle(0.04cm);
    \draw [fill=gray!30](1.35,0.7)circle(0.04cm);
    \draw [fill=gray!30](1.65,0.7)circle(0.04cm);

    \draw [draw=black, fill=gray!70, opacity=0.5] (0.15,0.3) rectangle (1.65, 0.7);
    \draw [draw=black, fill=yellow!70, opacity=0.3] (0,0) rectangle (1.8, 1);
    \node [color=black] at (0.9,0.5){\tiny Convex hull region  };

    \draw [<->, color=blue] (0.75,0.8) -- (1.05,0.8);
    \draw [<->, color=blue] (1.75,0.3) -- (1.75,0.7);      
    \node [color=black] at (1.85,0.5){\tiny $20$\,m  };

   \node [color=black] at (0.9,0.1){\tiny IOO scenario  };

   \end{tikzpicture}
    \end{subfigure}
  \end{center}
    \caption{Simulation results for 3GPP scenarios showing performance curves of various cases. The $90\%$ performances are shown in brackets inside the legend box.}
    \label{fig:sim}
  \end{figure}

In Fig.\ref{fig:sim}, different curves correspond to different cases for evaluation. The $90\%$ performances are shown in brackets inside the legends of the figure. Many cases are evaluated in the IOO scenario. Following observations can be drawn from the evaluations in IOO scenario
\begin{enumerate}
\item Larger bandwidth in FR2 results in better performance.
\item Multi-RTT has higher accuracy than TDOA based methods with signals being transmitted in DL and UL. It also relaxes requirements on network time synchronization.
\item Performance inside the convex-hull region of the IOO deployment is better than outside convex hull region. This shows importance of deployment in positioning. Maximizing the convex hull region while deploying base stations would improve the positioning performance.
  \item Due to comb-12, twelve orthogonal DL-PRS avoids interference in IOO scenario with twelve TRPs as shown in Fig.\ref{fig:sim}. Hence no performance difference is observed between interference-free and interference cases in IOO. In UMa and UMi scenarios, DL-PRS from multiple TRPs do interfere in interference case as number of orthogonal DL-PRS is twelve and number of TRPs is twenty-one.   
 \item  $8\times 8$ angle based positioning corresponds to the case with $64$ antenna elements to estimate angle of arrival in UL. Angle based positioning accuracy improves with number of antenna elements.  
 \end{enumerate}
  These evaluations are under the deployment and parameter assumptions discussed above and shown in Fig.\ref{fig:sim}. Better or worse accuracy can be obtained by different set of these selections.  

 \section{Conclusions}
3GPP Rel 16 provides fundamentals for 5G positioning. In this paper, we present the standardized 5G positioning architecture, procedures, signals and measurements. To illustrate the ability of the standardized components, we provide simulation results based on agreed key scenarios. with certain listed assumptions. The evaluations shows an accuracy technology potential of a few meters outdoors and a few deciemeters indoors, which makes 5G positioning highly applicable to many different use cases.
 

\bibliographystyle{IEEEtran} 
\bibliography{ref}
  \vspace*{-3\baselineskip}
 \begin{IEEEbiography}[{\includegraphics[width=1in,height=1.25in,clip,keepaspectratio]{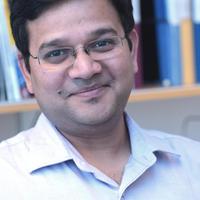}}]{Satyam Dwivedi}(satyam.dwivedi@ericsson.com) is a senior researcher at Ericsson research and team leader for positioning research and standardization team. He also works on propagation. He held position as researcher at the Royal Institute of Technology (KTH) in Stockholm, Sweden. He holds Master’s and  PhD from the Indian Institute of Science, Bangalore, India.  
  \end{IEEEbiography}
  \vspace*{-3\baselineskip}
  \begin{IEEEbiography}[{\includegraphics[width=1in,height=1.25in,clip,keepaspectratio]{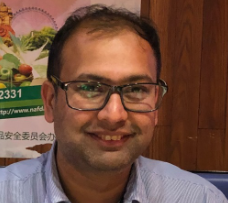}}]{Ritesh Shreevastav}(ritesh.shreevastav@ericsson.com) is a senior researcher and 3gpp RAN2 delegate at Ericsson research. He received MSc. in Telecommunications from Queens University Belfast, UK (2006) and Research Masters in Computer Science from Trinity College Dublin, Ireland (2008). His interests are Positioning, Machine Type Communications and Coverage enhancements.    
\end{IEEEbiography}
  \vspace*{-3\baselineskip}
\begin{IEEEbiography}[{\includegraphics[width=1in,height=1.25in,clip,keepaspectratio,angle=270]{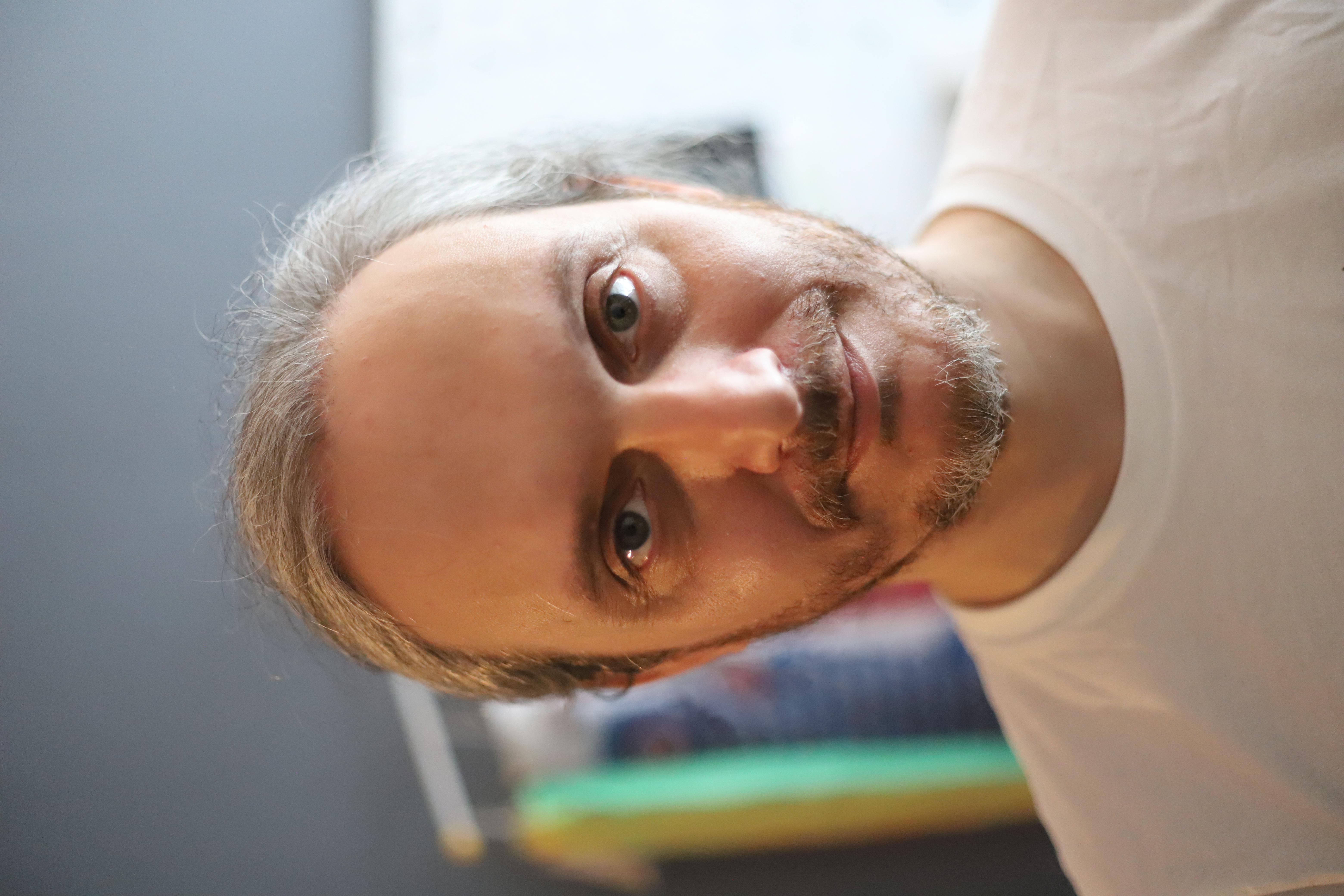}}]{Florent Munier} (florent.munier@ericsson.com) is a senior researcher at Ericsson research involved in 3GPP NR RAN1 standardization. He received a MSc in electrical and communication engineering from Napier University in Edinburgh, scotland in 2001, and a PhD in communication systems from Chalmers University of technology, Göteborg, Sweden in 2007. His current research interests are NR standardization of positioning and broadcasting. 
\end{IEEEbiography}
  \vspace*{-3\baselineskip}
\begin{IEEEbiography}[{\includegraphics[width=1in,height=1.25in,clip,keepaspectratio]{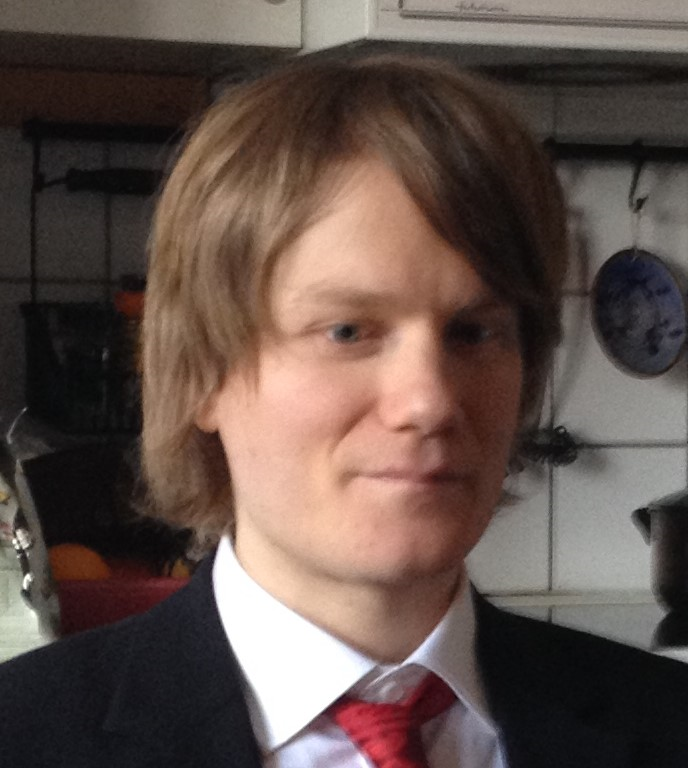}}]{Johannes Nygren}(johannes.nygren@ericsson.com) is a senior researcher at Ericsson research. He defended his PhD on Networked Control Systems at Uppsala University in 2016. His  research interests include robust estimation, channel estimation for positioning, and sensor fusion.
\end{IEEEbiography}
  \vspace*{-3\baselineskip}
  \begin{IEEEbiography}[{\includegraphics[width=1in,height=1.25in,clip,keepaspectratio]{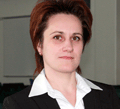}}]{Iana Siomina}(iana.siomina@ericsson.com) is currently Senior Specialist in RRM Performance, Ericsson, Sweden. She has been working at Ericsson Research since 2006, and is a 3GPP delegate of RRM and positioning. She has Ph.D. in Intra Informatics from Linköping University (2007), M.Sc. in Mathematics from KTH Royal Institute of Technology (2002), and M.Sc. in Computer Science from Stockholm University (2004).    
\end{IEEEbiography}
  \vspace*{-3\baselineskip}
  \begin{IEEEbiography}[{\includegraphics[width=1in,height=1.25in,clip,keepaspectratio]{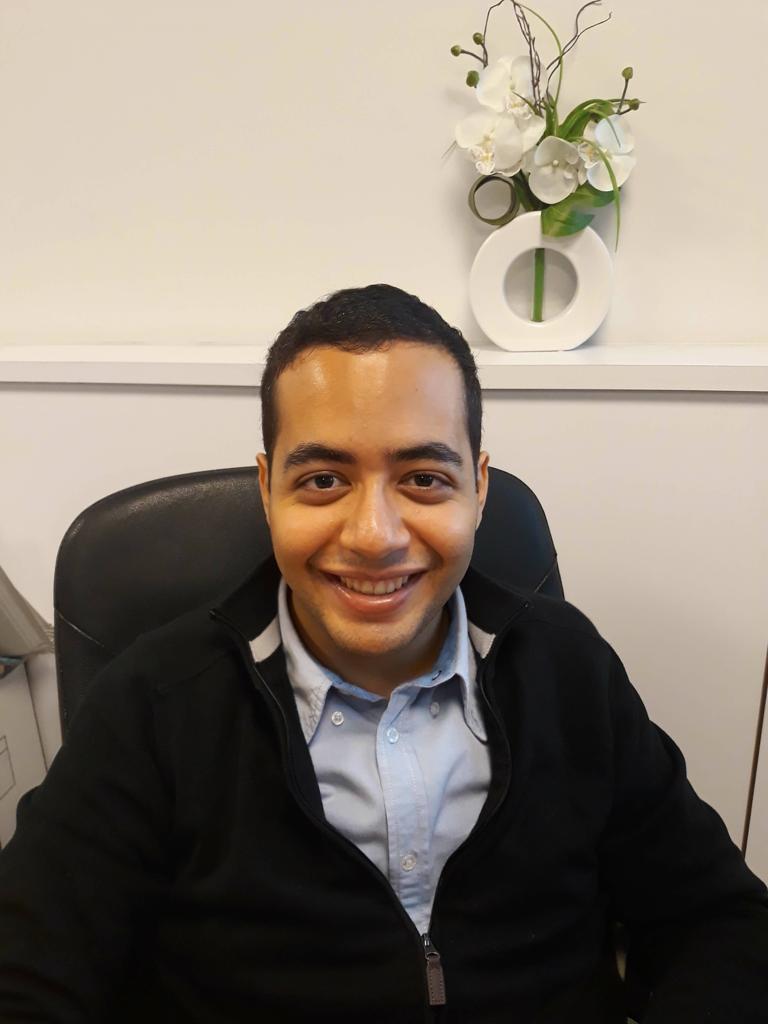}}]{Yazid Lyazidi} (yazid.lyazidi@ericsson.com) received his Master degree from Centrale Supelec France in 2014 and his PhD degree from UPMC Sorbonne Universities Paris in 2017.
      He is a standardization researcher at Ericsson Sweden.
\end{IEEEbiography}
  \vspace*{-3\baselineskip}
  \begin{IEEEbiography}[{\includegraphics[width=1in,height=1.25in,clip,keepaspectratio]{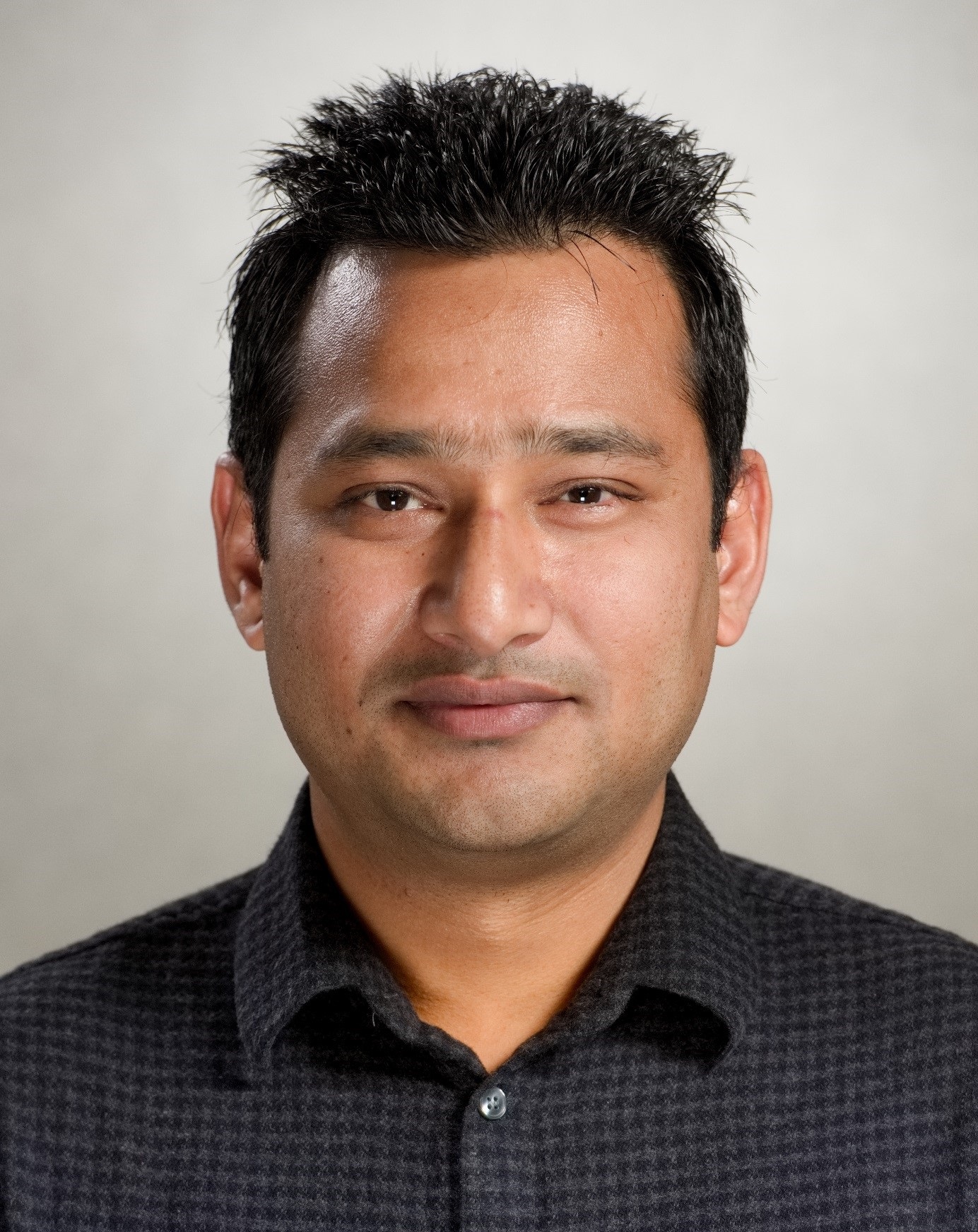}}]{Deep Shrestha}(deep.shreshtha@ericsson.com) is a senior researcher at Ericsson research, Linköping, Sweden since 2018. Deep is involved in positioning related topics for 5G standardisation and developing sensing and localization concepts for B5G Radio Access Technology. Deep did PhD in Signal Theory and Communications at Universitat Politècnica de Catalunya (UPC).    
\end{IEEEbiography}
  \vspace*{-3\baselineskip}
  \begin{IEEEbiography}[{\includegraphics[width=1in,height=1.25in,clip,keepaspectratio]{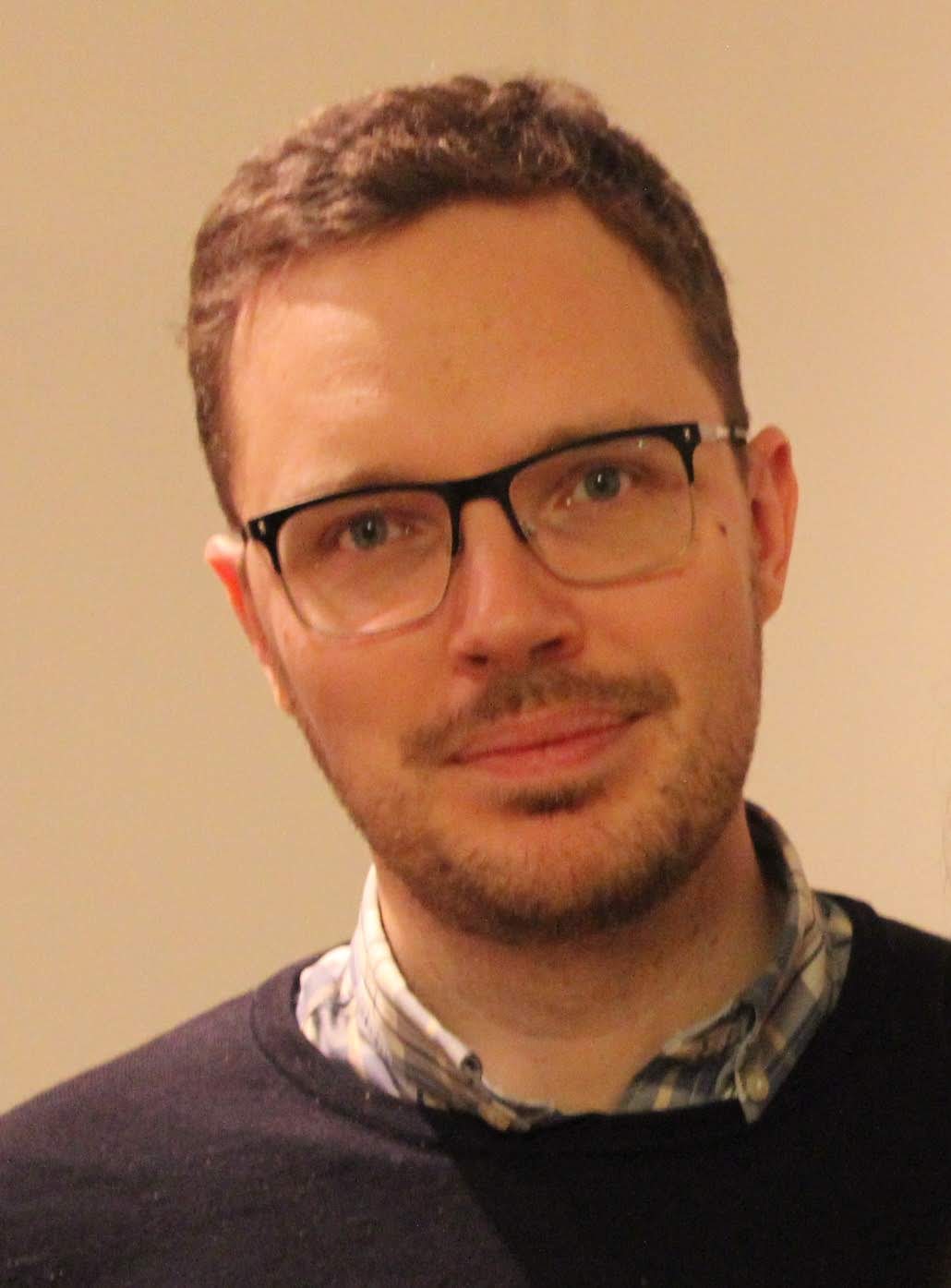}}]{Gustav Lindmark}(gustav.lindmark@ericsson.com) is a researcher at Ericsson research working with 3GPP standardization for positioning.
In 2020 he received PhD degree in Electrical Engineering from University of Linköping. He previously worked with software
development in Automotive industry and at Ericsson.
\end{IEEEbiography}
  \vspace*{-3\baselineskip}
\begin{IEEEbiography}[{\includegraphics[width=1in,height=1.25in,clip,keepaspectratio]{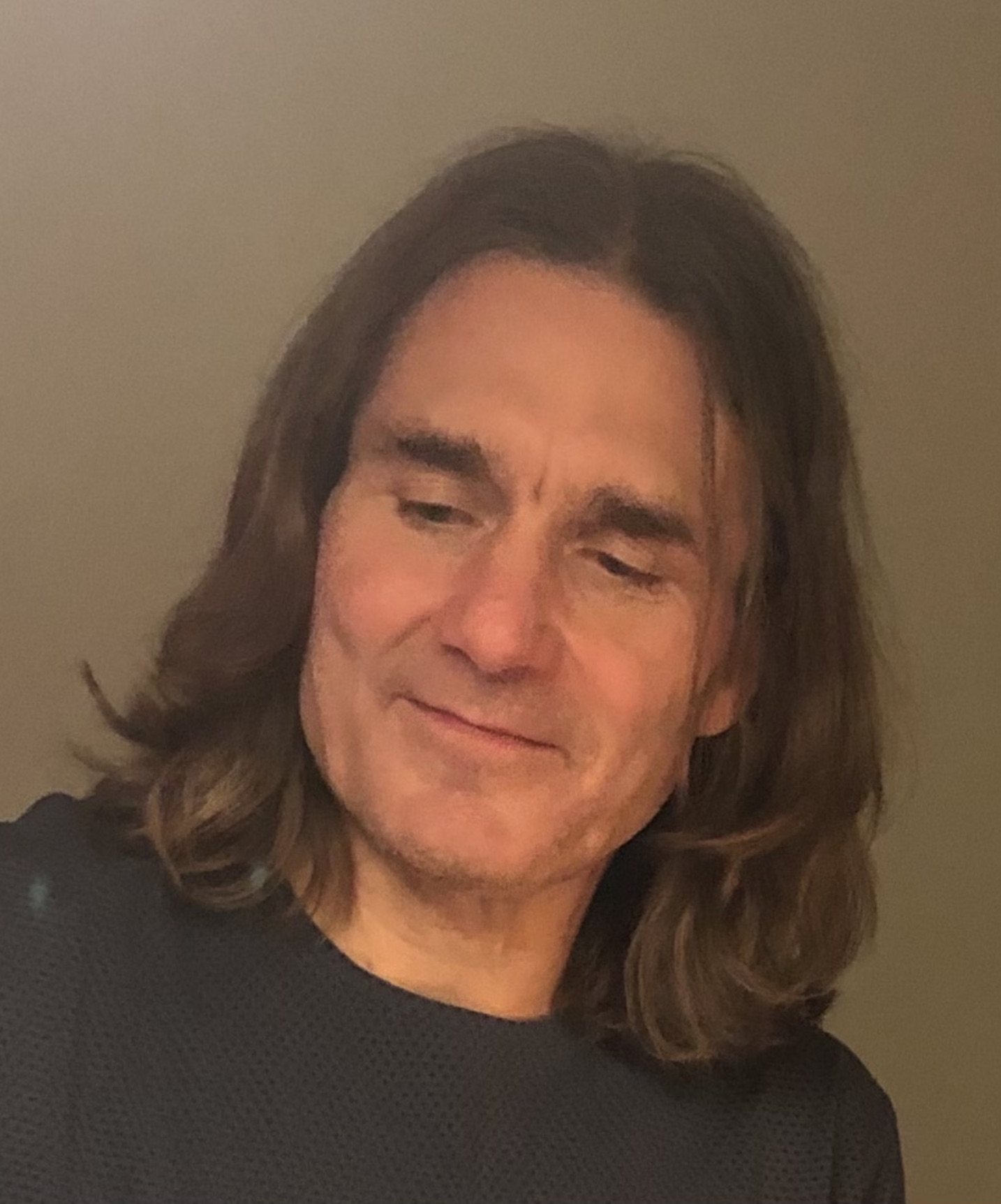}}]{Per Ernstr\"{o}m}(per.ernstrom@ericsson.com) is principal researcher at Ericsson AB focusing on 5G positioning. He was manager for the Ericsson 3GPP RAN standardization program 2010-2016. He received M.S. from KTH royal Institute of technology in 1989 and Ph.D. in theoretical particle physics from Stockholm University 1994 and had a research fellowship at the Nordic center for theoretical physic at the Niels Bohr Institute in Copenhagen 1994 to 1996.
\end{IEEEbiography}
\begin{IEEEbiography}[{\includegraphics[width=1in,height=1.25in,clip,keepaspectratio]{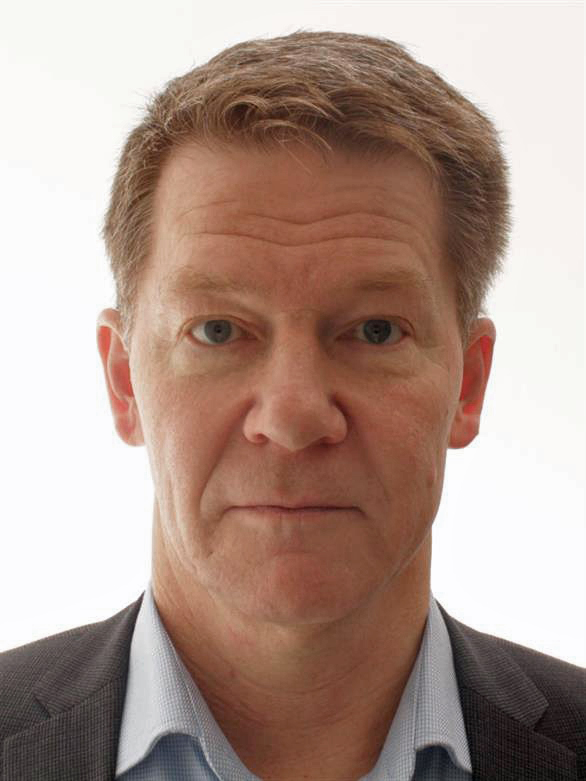}}]{Erik Stare}(erik.stare@ericsson.com) holds a M.Sc. degree (1984) in telecommunication from KTH (Royal Institute of Technology), Sweden. Between 1987-1992 (at Swedish Telecom) and 1992-2018 (at Teracom) he was deeply involved in developing systems and standards for digital terrestrial broadcasting. Since 2018 he is with Ericsson research holding Master Researcher position, working on positioning and a team leader for 5G/NR multicast/broadcast standardization while being a 3GPP RAN1 delegate.
\end{IEEEbiography}
  \vspace*{-3\baselineskip}
  \begin{IEEEbiography}[{\includegraphics[width=1in,height=1.25in,clip,keepaspectratio]{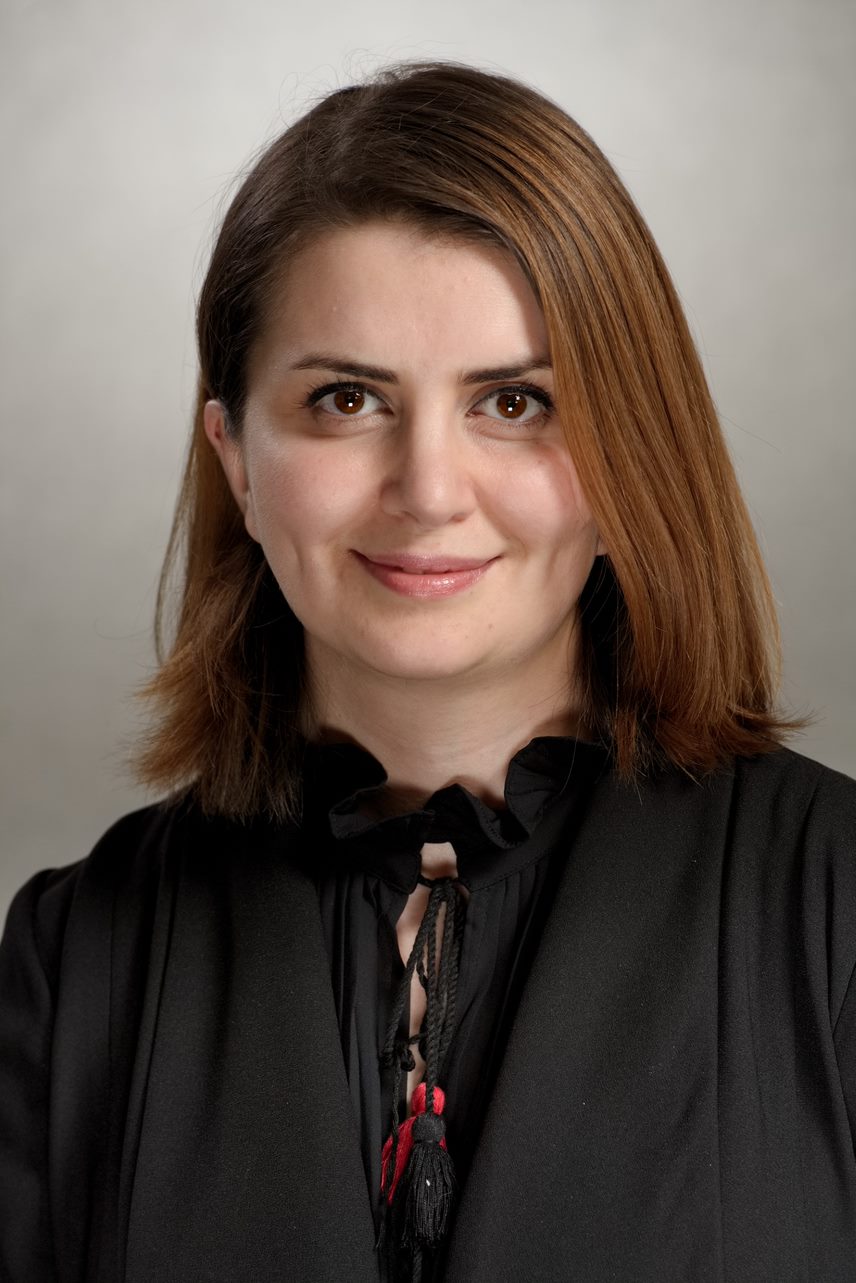}}]{Sara Modarres Razavi}(sara.modarres.razavi@ericsson.com)(PhD)  is  a  senior  researcher at Ericsson research involved in 3GPP positioning standardization since Rel.13. She is currently the project manager of LTE and NR 3GPP standardization project. She holds MSc in Hardware for Wireless Communication from Chalmers University of Technology (2008) and PhD in Infra-Informatics from Linköping University (2014) in Sweden.
  \end{IEEEbiography}
  \vspace*{-3\baselineskip}
  \begin{IEEEbiography}[{\includegraphics[width=1in,height=1.25in,clip,keepaspectratio]{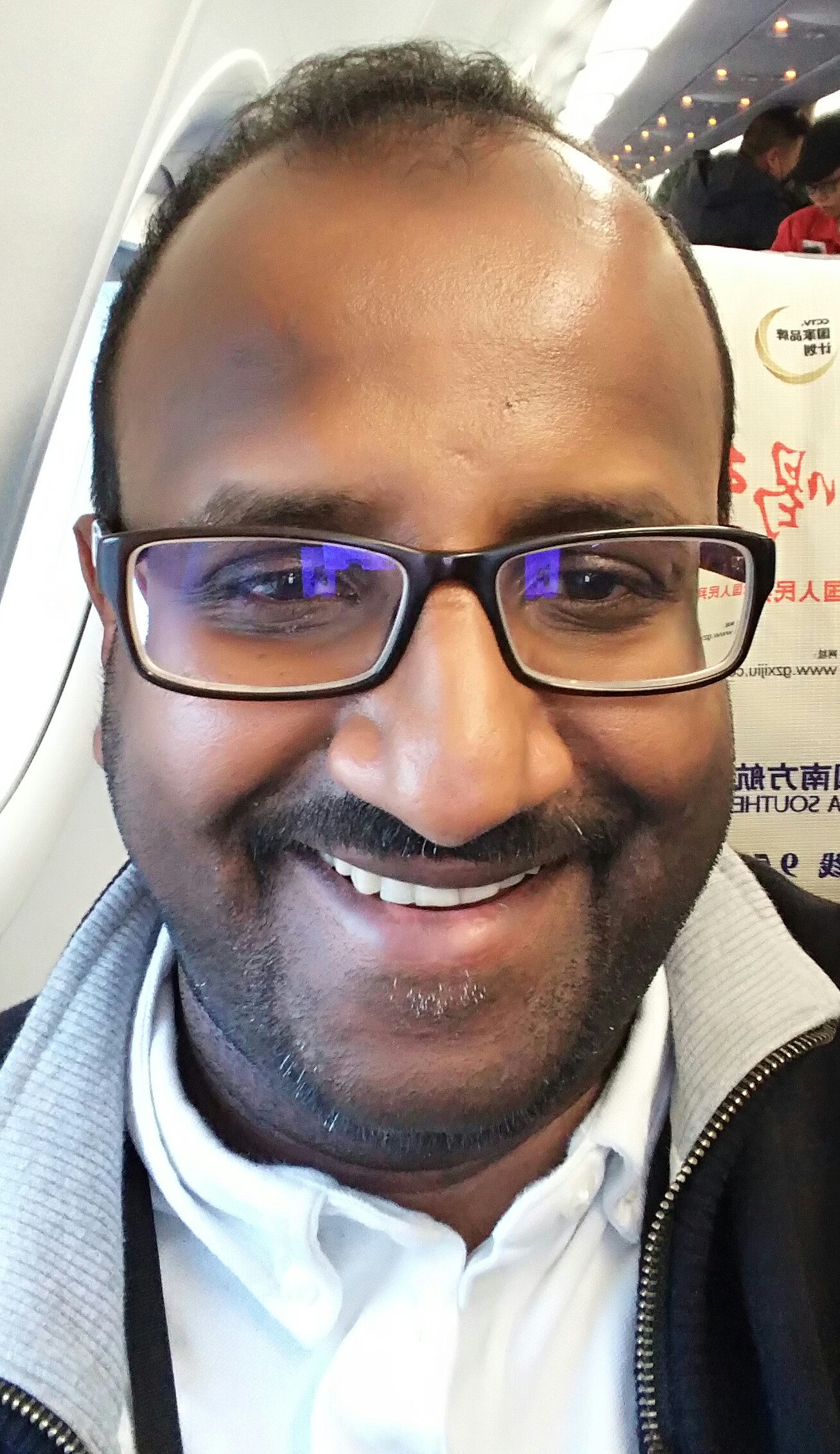}}]{Siva Muruganathan}(siva.muruganathan@ericsson.com) is currently a researcher and 3GPP RAN1 delegate with Ericsson Canada, Ottawa, Ontario. He previously held research/postdoctoral positions at BlackBerry Limited, CRC Canada, and the University of Alberta, Canada. His recent standardization work is in the areas of MIMO, positioning and air-to-ground communications.
      \end{IEEEbiography}
  \vspace*{-3\baselineskip}
  \begin{IEEEbiography}[{\includegraphics[width=1in,height=1.25in,clip,keepaspectratio]{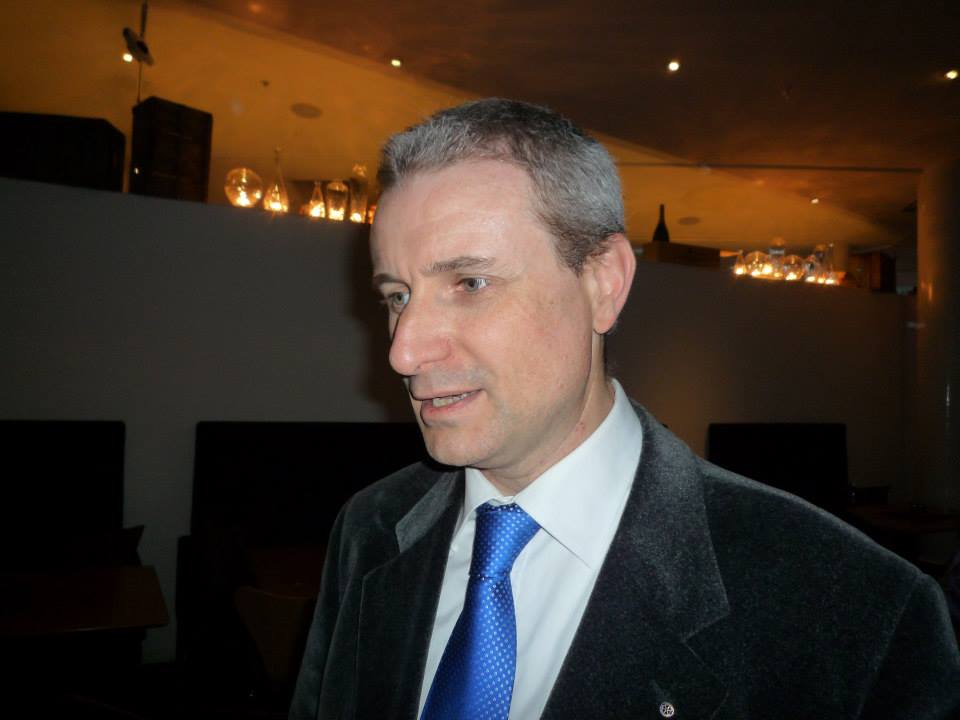}}]{Gino Masini} (gino.masini@ericsson.com) is principal researcher with Ericsson in Sweden. He received his MSc in Electronics Engineering from Politecnico di Milano in 1996, and his MBA from SDA Bocconi School of Management in Milano in 2008.
    He joined Ericsson in 1999, working with microwave radio links and MMIC development. Since 2009 he has worked with 4G and 5G RAN architecture and is currently in his second term as 3GPP RAN WG3 Chairman.
\end{IEEEbiography}
  \vspace*{-3\baselineskip}
\begin{IEEEbiography}[{\includegraphics[width=1in,height=1.25in,clip,keepaspectratio]{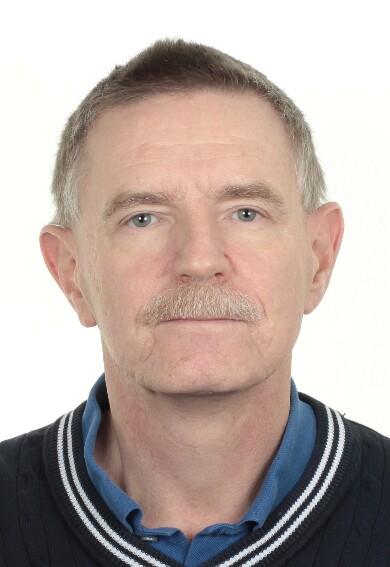}}]{Åke Busin}(ake.busin@ericsson.com)
received his M.Sc. degree electrical engineering from KTH Royal Institute of Technology in Stockholm, in 1984.  He is currently employed with Ericsson AB, Stockholm, where he is working as a Senior Specialist Location Based Services.
\end{IEEEbiography}
  \vspace*{-3\baselineskip}
  \begin{IEEEbiography}[{\includegraphics[width=1in,height=1.25in,clip,keepaspectratio]{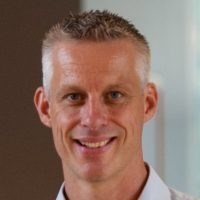}}]{Fredrik Gunnarsson}(fredrik.gunnarsson@ericsson.com)
is an expert in positioning and RAN automation at Ericsson research. He works on research, concepts, standardization, prototyping,
use case adaptation and realization aspects of RAN automation and positioning. He also holds a part-time position as adjunct professor at Linköping University, Sweden.  He obtained his MSc and PhD in Electrical Engineering from Linköping University in 1996 and 2000 respectively.

  \end{IEEEbiography}

  \end{document}